\documentclass[prb,twocolumn,aps,a4paper,showpacs]{revtex4}
\usepackage{dcolumn}
\usepackage{amsmath}
\usepackage{graphicx}
\usepackage{latexsym}
\usepackage{amsfonts}
\usepackage{amssymb}

\begin{document}

\def\tg{\mbox{\textsl{g}}}
\def\bS{\mbox{\boldmath $\Sigma$}}
\def\bDelta{\mbox{\boldmath $\Delta$}}
\def\bcalE{\mbox{\boldmath ${\cal E}$}}
\def\bcalG{\mbox{\boldmath ${\cal G}$}}
\def\bG{\mbox{\boldmath $G$}}
\def\bSS{\mbox{\boldmath $S$}}
\def\bK{\mbox{\boldmath $K$}}
\def\bT{\mbox{\boldmath $T$}}
\def\bU{\mbox{\boldmath $U$}}
\def\bV{\mbox{\boldmath $V$}}
\def\bg{\mbox{\boldmath $g$}}
\def\gC{\mbox{\boldmath $C$}}
\def\gZ{\mbox{\boldmath $Z$}}
\def\gR{\mbox{\boldmath $R$}}
\def\gN{\mbox{\boldmath $N$}}
\def\ua{\uparrow}
\def\da{\downarrow}
\def\a{\alpha}
\def\b{\beta}
\def\g{\gamma}
\def\G{\Gamma}
\def\d{\delta}
\def\D{\Delta}
\def\e{\epsilon}
\def\ve{\varepsilon}
\def\z{\zeta}
\def\h{\eta}
\def\th{\theta}
\def\k{\kappa}
\def\l{\lambda}
\def\L{\Lambda}
\def\m{\mu}
\def\n{\nu}
\def\x{\xi}
\def\X{\Xi}
\def\p{\pi}
\def\P{\Pi}
\def\r{\rho}
\def\s{\sigma}
\def\S{\Sigma}
\def\t{\tau}
\def\f{\phi}
\def\vf{\varphi}
\def\F{\Phi}
\def\c{\chi}
\def\w{\omega}
\def\W{\Omega}
\def\Q{\Psi}
\def\q{\psi}
\def\de{\partial}
\def\inf{\infty}
\def\ra{\rightarrow}
\def\bra{\langle}
\def\ket{\rangle}

\title{Time-Dependent Partition-Free Approach in Resonant Tunneling Systems}

\author{Gianluca Stefanucci and Carl-Olof Almbladh}

\affiliation{Department of Solid State Theory, Institute of Physics, 
Lund University\\
S\"olvegatan 14 A, 223 62 Lund, Sweden}

\date{\today}

\begin{abstract}

An extended Keldysh formalism, well suited to properly take into 
account the initial correlations, is used  
in order to deal with the time-dependent current response of a
resonant tunneling system. 
We use a \textit{partition-free} approach by Cini in which
the whole system is in equilibrium before an external bias
is switched on.
No fictitious partitions 
are used. Despite a more involved formulation, 
this partition-free approach has many appealing features being much 
closer to what is experimentally done. In particular, besides the 
steady-state responses one can also calculate physical dynamical responses.
In the noninteracting case we clarify under what circumstances a 
steady-state current develops and compare our result with the one 
obtained in the partitioned scheme. We prove a Theorem of asymptotic 
Equivalence between the two schemes for arbitrary time-dependent 
disturbances. We also show that the steady-state current is 
independent of the history of the external perturbation 
(Memory Loss Theorem). In the so called wide-band limit
an analytic result for the time-dependent current is obtained. 
In the interacting case we work out the lesser 
Green function in terms of the self energy and we recover a well known 
result in the long-time limit. In order to overcome the complications 
arising from a self energy which is nonlocal in time we propose an   
exact non-equilibrium Green function approach based on Time Dependent Density 
Functional Theory. The equations are no 
more difficult than an ordinary Mean Field treatment. We show
how the scattering-state scheme by Lang follows from our formulation.
An exact formula for the steady-state current of an arbitrary 
interacting resonant tunneling system is obtained.
As an example the time-dependent current response 
is calculated in the Random Phase Approximation.

\end{abstract}


\maketitle

\section{Introduction}
\label{intro}

A resonant tunneling system is essentially a mesoscopic region, 
typically a semiconductor heterostructure, coupled to 
two metallic leads, which play the role of charge reservoirs. 
In a real experiment the whole system is in thermodynamic 
equilibrium before the external disturbance is switched on and one can assign a 
unique temperature $\b^{-1}$ and chemical potential $\m$. Therefore, the initial 
density matrix is $\r\sim \exp[-\b(H-\m N)]$ where $H$ is the total 
Hamiltonian and $N$ is the total number of particles. By applying a 
bias to the leads at a given time, charged particles will start to flow 
through the central device from one lead to the other. As far as the leads 
are treated as \textit{noninteracting}, it is not obvious that in the 
long-time limit a steady-state current can ever develop. The reason 
behind the uncertainty is that the bias 
represents a large perturbation and, in the absence of dissipative 
effects, \textit{e.g.}, electron-electron or electron-phonon scatterings, 
the return of time-translational invariance is not granted. 

An alternative approach 
to this quantum transport problem has been suggested by Caroli \textit{
et al.}\cite{caroli1,caroli2} who state: 
``It is usually considered that a description of the system as 
a whole does not permit the calculation of the current''.\cite{caroli1}  
Their approach is based on a fictitious \textit{partition} where 
the left and right leads are treated as two isolated subsystems in the 
remote past. Then, one can fix a chemical potential $\m_{\a}$ and 
a temperature $\b^{-1}_{\a}$ for each lead, $\a=L,R$. In this picture the 
initial density matrix is given by $\r\sim \exp[-\b_{L}(H_{L}-\m_{L} 
N_{L})]\exp[-\b_{R}(H_{R}-\m_{R} N_{R})]$, where $H_{L,R}$ and $N_{L,R}$ 
now refer to the isolated $L,R$ lead. The current will flow through 
the system once the contacts between the device and the leads have  
been established. Hence, the time-dependent perturbation is a lead-device 
hopping rather than a local one-particle level-shift. Since the 
device is a mesoscopic object, it is reasonable to assume that the 
hopping perturbation does not alter the thermal equilibrium of the 
left and right charge reservoir and that 
a non-equilibrium steady state will eventually be reached. This argument is very 
strong and remains valid even for \textit{noninteracting} leads. 
Actually, the partitioned approach by Caroli \textit{et al.} was originally applied to 
a tight-binding model\cite{caroli1} describing a 
metal-insulator-metal tunneling junction and then extended to the 
case of free electrons subjected to an arbitrary 
one-body potential.\cite{caroli2} This extension was questioned by 
Feuchtwang;\cite{feuchtwang1,feuchtwang2}  
the controversy was about the appropriate 
choice of boundary conditions for the uncontacted-system Green 
functions. In later years the non-equilibrium Green function 
techniques\cite{kb-book,keldysh-jetp1965} in the partitioned 
approach framework were mainly applied to investigate steady-state 
situations. An important breakthrough in time-dependent 
non-equilibrium transport was achieved by Wingreen  
\textit{et al.}\cite{wingreen,jauho,jauhocond,haug}  
Still in the framework of the partitioned approach, 
they derive an expression for the fully nonlinear, 
time-dependent current in terms of the Green functions 
of the mesoscopic region (this embedding procedure holds only for 
noninteracting leads). 
Under the physical assumption that the initial correlations 
are washed out in the long-time limit, their formula is well 
suited to study the response to external time-dependent voltages 
and contacts. 

The limitations of the partitioned approach are essentially three. First,
it is difficult to partition the electron-electron interactions between the 
leads and between the leads and the device. These interactions are 
important for establishing dipole layers and charge transfers 
which shape the potential landscape in the device region. 
Second, there is a crucial assumption of equivalence 
between the long-time behavior of the 1) initially partitioned and 
biased system once the \textit{coupling} between the subsystems is  
established and 2) the whole partition-free system in thermal 
equilibrium once the \textit{bias} is established. Third, 
the transient current has no direct physical interpretation 
since in a real experiment one switches on the bias and 
not the contacts; moreover,  
there is no well defined prescription which fixes the initial 
equilibrium distribution of the isolated central device. 

In this paper we use a partition-free scheme without the 
above limitations. This conceptually different time-dependent approach 
has been proposed by Cini.\cite{cini} He developed the general 
theory for the case of free electrons described in terms of a discrete 
set of states and a continuum set of states with focus on 
semiconductor junction devices. For a one-dimensional free-electron 
system subjected to a time-dependent perturbation of the form 
$U\Theta(t)\Theta(x)$, where $U$ is the applied bias and $(x,t)$ 
is the space-time variable, the Cini theory yields a current-voltage 
characteristics which agrees with the one obtained by 
Feuchtwang\cite{feuchtwang1,feuchtwang2} in the partitioned approach. 
This result is particularly important since it shows that a steady state 
in a partition-free scheme develops even in the 
noninteracting case. Moreover, it demonstrates an equivalence which 
had previously been assumed. In the present work we extend the 
partition-free approach to noninteracting resonant tunneling systems 
and also to \textit{interacting} such systems - in both cases using  
arbitrary time-dependent disturbances. We shall clarify under 
what circumstances a non-equilibrium steady state can develop and 
discuss the equivalence of the current-voltage characteristics 
obtained by Jauho \textit{et al.}\cite{jauho} and that obtained by us. 
One of the advantages of the partition-free scheme over the 
traditional methods lies in the ability of the former to calculate 
transient physical (\textit{i.e.} measurable) current responses.

The plan of the paper is the following. In Section \ref{keldyshth} we 
develop the general formalism which properly accounts for the initial 
correlations. We derive a solution of the Keldysh equations for the 
lesser and the greater Green functions in noninteracting and interacting 
systems. An exact and alternative treatment based on Time Dependent Density 
Functional Theory\cite{rg} (TDDFT) is proposed in order to calculate the 
total nonlinear time-dependent current. The current response of a 
noninteracting resonant tunneling system is discussed in Section 
\ref{model}. We specify when the partitioned and 
the partition-free schemes yield the same asymptotic current (Theorem 
of Equivalence) and how this current  
may depend on history (Memory Loss Theorem). 
The general results are illustrated by model calculations. 
In Section \ref{interacting} we consider an interacting resonant tunneling 
system with interacting leads. The TDDFT approach is compared with 
earlier works by Lang \textit{et al.}\cite{Lang1,Lang2} and Taylor 
\textit{et al.}\cite{Taylor1,Taylor2} Assuming that a steady state 
is reached we write down an exact formula for the nonlinear 
steady-state current. As a simple example we also study the current 
response in the Random Phase Approximation (RPA) of a 
capacitor-device-capacitor junction. Our main conclusions are 
summarized in Section \ref{conclusion}.

\section{General Formulation}
\label{keldyshth}

\subsection{Noninteracting Systems in the Presence of an External 
Disturbance}
\label{nonint}

Let us consider a system of noninteracting electrons 
described by an unperturbed Hamiltonian 
\begin{equation}
H_{0}=\sum_{mn}T_{m,n}c_{m}^{\dag}c_{n},
\quad\quad (\bT)_{m,n}=T_{m,n}
\label{eq:h0}
\end{equation}
and by a time-dependent disturbance of the form
\begin{equation}
H_{U}(t)=\sum_{mn}U_{m,n}(t)c^{\dag}_{m}c_{n},
\quad [\bU(t)]_{m,n}=U_{m,n}(t),
\label{eq:hu}
\end{equation}
with $\bU(t)=0$ for any $t\leq t_{0}$.
In Eqs. (\ref{eq:h0})-(\ref{eq:hu}), $c_m, c^\dagger_n$ are
Fermi operators in some suitable basis, and 
we use boldface to indicate matrices in one-electron
labels. 
Without loss of generality one can 
take $t_{0}=0$. The system is in equilibrium for negative times.

\subsubsection{Elementary Derivation}

We first obtain the Green function by elementary means without
resorting to any Keldysh techniques. For a noninteracting
system everything is known once we know how to propagate 
the one-electron orbitals in time and how they are populated
before the system is perturbed. The time evolution is fully
described by the retarded or advanced Green functions $\bG^{\rm R,A}$,
and the initial population 
at zero time, \textit{ i.e.}, by $\bG^<(0;0)$. 
The real-time Green functions are defined by
$$
\bG^{\rm R,A}(t;t^\prime) = \pm  
\Theta(\pm t \mp t^\prime)\left[
\bG^{>}(t;t^\prime)-\bG^{<}(t;t^\prime)
\right],
$$
with
\begin{eqnarray}
G^>_{m,n}(t;t^\prime) &=& -i \langle c_m(t) c^\dagger_n(t^\prime) \rangle ,
\nonumber \\  
G^<_{m,n}(t;t^\prime) &=& i \langle c^\dagger_n (t^\prime) c_m(t) \rangle,
\nonumber 
\end{eqnarray}
where the operators are Heisenberg operators and where the averages
are with respect to the equilibrium grand-canonical ensemble.
Because there are no inter-particle interactions, the equation of motion
for the electron operators simplifies to
\begin{displaymath}
i \dot{c}_m(t) = \sum_n K_{m,n}(t)c_n(t) 
\end{displaymath}
where $\bK(t)\equiv \bT + \bU(t)$ 
is the full one-body Hamiltonian matrix. 
Consequently, the time evolution of $c_m$ is given by the
one-electron evolution matrix $\bSS(t)$,
$c_m(t) = \sum_n S_{m,n}(t) c_n(0)$, 
where $\bSS$ obeys 
$i\dot{\bSS}(t)=\bK(t)\bSS(t)$,
with initial value $\bSS(0)$ = 1. We insert the time-evolved operators
in the definitions of the $\bG$ matrices to obtain
\begin{equation}
\bG^{\rm R,A}(t;t')=\mp i
\Theta(\pm t \mp t^\prime)
\bSS(t)\bSS^{\dag}(t'),
\label{gra} 
\end{equation}
and
\begin{eqnarray}
\bG^{\lessgtr}(t; t') &=& \bSS(t) \bG^{\lessgtr}(0;0)\bSS^{\dag}(t')
\nonumber \\
 &=& \bG^{\rm R}(t;0) \bG^{\lessgtr}(0;0)\bG^{\rm A}(0;t'), 
 \label{g<}
\end{eqnarray}
where the last equality holds for any $t,t'>0$. 
We observe that the instantaneous current can be expressed in terms of 
$\bG^{<}(t; t)$,
and thus the problem of finding the current is reduced to 
that of finding the retarded Green function and the equilibrium
population of the one-electron levels. We note in passing that
the initial populations can be expressed as 
$\bG^{<}(0;0)=i f(\bT)$, 
where $f$ is the Fermi function.
Because $\bT$ is a matrix, so is $f(\bT)$.

The above solution for the lesser/greater Green function 
was also derived by Cini\cite{cini} with an 
equation-of-motion 
approach. He also pointed out that they can be derived in the 
framework of the Keldysh formalism\cite{keldysh-jetp1965} as a 
finite-temperature extension of a treatment by Blandin \textit{ et 
al.}\cite{blandin} 

\subsubsection{Derivation Based on the Keldysh Technique}

In this Subsection we give an alternative derivation of 
Eq. (\ref{g<}) using an extension of the Keldysh 
formalism. There are two reasons for giving another derivation. On 
one hand, we will use the Keldysh formalism taking due account of the 
prescribed integration along the imaginary axis. This will allow us to 
understand what kind of approximations are made in the partitioned 
approach. On the other hand, the derivation below clearly shows how 
the electron-electron interaction can be included.

\begin{figure}[htbp]
\begin{center}
\includegraphics*[scale=0.8]{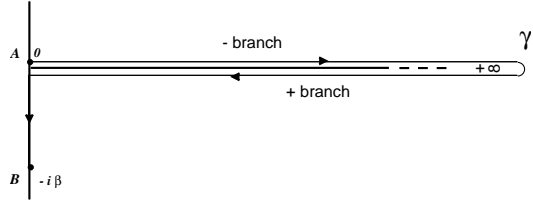}
\caption{{\footnotesize Contour suited to include the effect of the 
initial correlations, see also Section \ref{intkel}.}}
\label{transcontour}
\end{center}
\end{figure}

We introduce the Green function
\begin{equation}
G_{m,n}(z;z')= -i\bra {\cal T}
[c_{m}(z)\bar{c}_{n}(z')]
\ket
\label{gf}
\end{equation}
which is path-ordered 
on the oriented contour $\g$ of Fig. \ref{transcontour}. In Eq. (\ref{gf}) 
$z=t+\t$ is the complex variable running on $\g$ with 
$t=\Re[z]$, $\t=i\,\Im[z]$ while $A$ and $B$ are the 
end-points of $\g$. Further, $c_m(z)$ and $\bar{c}_n(z)$
are Heisenberg operators defined by the non-unitary 
evolution operator for complex times $z$. They are 
in general not Hermitian conjugates of one another, but the
usual equal-time anticommutation relations 
$\{ c_m(z), \bar{c}_n(z) \} = \delta_{m,n}$ 
are still obeyed. 
As before the average is the grand-canonical average. 
On the vertical track going from 0 to $-i\b$ we have 
$\bK(\t)=\bK(0)=\bT$ independent of $\t$. 
Therefore, the Green function satisfies the relations 
\begin{equation}
\begin{array}{l}
\bG(A;z')=-{\rm e}^{\b\m}\bG(B;z')\\
\bG(z;A)=-{\rm e}^{-\b\m}\bG(z;B)
\end{array}.
\label{mrg<>}
\end{equation}

Next, we write the total Hamiltonian $H(t)=H_{0}+H_{U}(t)$ as the 
sum of a diagonal term and an off-diagonal one
$$
H(t)=\sum_{m}\ve_{m}(t)c^{\dag}_{m}c_{m}+
\sum_{mn}V_{m,n}(t)c^{\dag}_{m}c_{n}.
$$
The quantities $\ve_{m}(\t)=\ve_{m}$ and 
$V_{m,n}(\t)=V_{m,n}$ are constants on the vertical track. 
[The decomposition above is completely general. In our model examples 
discussed later, the diagonal part $[\bcalE(z)]_{m,n}=\d_{m,n}\ve_{m}(z)$ 
will represent an uncontacted system and the off-diagonal 
$[\bV(z)]_{m,n}=V_{m,n}(z)$ the contacts.]
The Green function $\bG(z;z')$ is obtained by solving the equation of motion    
\begin{equation}
\left\{i\frac{d}{dz}-\bcalE(z)-\bV(z)\right\}\bG(z;z')=\d(z-z'),
\label{eomg}
\end{equation}
[and its adjoint] with boundary conditions (\ref{mrg<>}). 
We define ${\bf g}(z;z')$ as the uncontacted Green function. The  
${\bf g}$ satisfies Eq. (\ref{eomg}) with $\bV=0$ and obeys the 
same boundary conditions of the contacted $\bG$. 
The unique ${\bf g}$ resulting from such a scheme belongs to 
the Keldysh space\cite{daniele} and can be written as 
$$
{\bf g}(z;z')=\Theta(z,z'){\bf g}^{>}(z;z')+\Theta(z',z){\bf g}^{<}(z;z'),
$$
where $\Theta(z,z')=1$ if $z$ is later than $z'$ on $\g$ 
and zero otherwise. ${\bf g}^{>}(z;z')$ is analytic for any $z$ later 
than $z'$ while ${\bf g}^{<}(z;z')$ is analytic for any $z'$ later 
than $z$; they are given by 
\begin{eqnarray}
{\bf g}^{<}(z;z')&=&if(\bcalE ){\rm e}^{-i\int_{z'}^{z}d\bar{z}\bcalE(\bar{z})},
\nonumber \\ 
{\bf g}^{>}(z;z')&=&i[f(\bcalE)-1]{\rm e}^{-i\int_{z'}^{z}d\bar{z}\bcalE(\bar{z})},
\label{lg<>}
\end{eqnarray}
where $\bcalE\equiv\bcalE(0)$ and the integral appearing in the 
exponential function is a contour integral along $\g$ going from $z'$ to $z$. 
Choosing $z$ and $z'$ on the real axis ${\bf g}^{<}$ and ${\bf g}^{>}$ 
reduce to the real-time lesser and greater component. From 
Eqs. (\ref{lg<>}) one can easily verify that
the corresponding retarded and advanced component can be written as
\begin{equation}
{\bf g}^{\rm R,A}(t;t')=\mp i\Theta(\pm t\mp t')
{\rm e}^{-i\int_{t'}^{t}d\bar{t}\bcalE(\bar{t})}.
\label{lgra}
\end{equation}

The uncontacted ${\bf g}$ allows to convert Eqs. (\ref{eomg})
into an integral equation which preserves the relations (\ref{mrg<>}):
\begin{equation}
\bG(z;z')={\bf g}(z;z')+\int_{\g}
d\bar{z}\;{\bf g}(z;\bar{z})\bV(\bar{z})\bG(\bar{z};z').
\end{equation}  
Using the Langreth theorem\cite{langreth} one finds
\begin{equation}
\bG^{\lessgtr}=\left[\d+
\bG^{\rm R}\cdot \bV\right]\cdot {\bf g}^{\lessgtr} +
\bG^{\lessgtr}\cdot \bV \cdot {\bf g}^{\rm A}
+\bG^{\rceil}\star \bV \star {\bf g}^{\lceil},
\label{newder1}
\end{equation}
where we have used the short hand notation $\cdot$ to denote 
integrals along the real axis, going from 0 to $\inf$, 
and $\star$ for integrals along the 
imaginary vertical track, going from 0 to $-i\b$. For the sake of 
clarity we have also introduced the symbols $\rceil$ and $\lceil$: any 
function with the superscript $\rceil$ is intended to have a real first 
argument and an imaginary second argument; the opposite is specified 
by $\lceil$. In Eq. (\ref{newder1}), $\bV(z;z')\equiv \d(z-z')\bV(z)$; for 
$\bV$ we don't need to say more since it is always foregone and 
followed by $\cdot$ or $\star$ so that no ambiguity arises. In 
particular we note that $\star \bV \star$ implies a simple matrix 
multiplication since along the vertical track 
$\bV$ is a constant matrix times the delta function.

The equation for $\bG^{\lessgtr}$ contains $\bG(t;\t)$ with 
one real and one imaginary argument. This coupling does not allow to get 
a closed equation for $\bG(t;t')$ with two real arguments, 
unless $\bV=0$ on the vertical track. Conversely, $\bG^{\rm R}$ and 
$\bG^{\rm A}$ satisfy an integral equation without any coupling:
\begin{equation}
\bG^{\rm R,A}={\bf g}^{\rm R,A}+
\bG^{\rm R,A}\cdot \bV\cdot {\bf g}^{\rm R,A}.
\label{newder2}
\end{equation}
Eq. (\ref{newder1}) can be solved for $\bG^{\lessgtr}$ and one obtains
\begin{eqnarray}
\bG^{\lessgtr}&=&\left[\d+
\bG^{\rm R}\cdot \bV
\right]\cdot {\bf g}^{\lessgtr}\cdot 
\left[
\d+\bV\cdot \bG^{\rm A}
\right]
\nonumber \\ &&+
\bG^{\rceil}\star \bV\star {\bf g}^{\lceil}\cdot 
\left[
\d+\bV\cdot \bG^{\rm A}
\right]\;.
\label{newder3}
\end{eqnarray}
From Eq. (\ref{lg<>}) and Eq. (\ref{lgra}) we have
${\bf g}^{\lessgtr}(t;t')={\bf g}^{\rm R}(t;0)
{\bf g}^{\lessgtr}(0;0){\bf g}^{\rm A}(0;t')$ and 
$ {\bf g}(\t;t)=-i{\bf g}(\t;0){\bf g}^{\rm A}(0;t)$, 
so that Eq. (\ref{newder3}) can be rewritten as
\begin{eqnarray}
\bG^{\lessgtr}(t;t')&=&\bG^{\rm R}(t;0){\bf g}^{\lessgtr}(0;0)
\bG^{\rm A}(0;t')\nonumber \\ && -i
\left[\bG^{\rceil}\star \bV\star {\bf g}\right](t;0) \bG^{\rm A}(0;t').
\label{newder4}
\end{eqnarray}
The above expression for $\bG^{\lessgtr}$ deserves a brief 
comment. Indeed, the first term on the r.h.s. is exactly what one got in the 
partitioned approach, where the hopping parameters $V_{m,n}$ vanish along the 
vertical track. It is usually argued that if 
$t,t'\ra\inf$ the second term vanishes. However, we point out that in the 
noninteracting case this is not true. If in the long-time 
limit some physical response functions, \textit{ e.g.}, the current, are 
correctly reproduced by using the partitioned $\bG^{\lessgtr}(t;t')=
\bG^{\rm R}(t;0){\bf g}^{\lessgtr}(0;0)\bG^{\rm A}(0;t')$ other 
kind of argumentations should be invoked. We shall come to 
this point later on.

To proceed further we need the Dyson equation for $\bG(t;\t)$. 
Exploiting the identity ${\bf g}(t;\t)=i{\bf g}^{\rm R}(t;0)
{\bf g}(0;\t)$, we find
\begin{equation}
\bG(t;\t)=i\bG^{\rm R}(t;0){\bf g}(0;\t)+
\left[\bG^{\rceil}\star \bV\star {\bf g}\right](t;\t).
\label{newder6}
\end{equation}
Eq. (\ref{newder6}) can be solved for $\bG(t;\t)$. 
From the Dyson equation $\bG(\t;\t')={\bf g}(\t;\t')+
\left[\bG\star \bV\star {\bf g}\right](\t;\t')$, 
it follows that $\left[
\d-\bV\star {\bf g}
\right]^{-1}(\bar{\t};\t)=
\left[\d+\bV\star \bG\right](\bar{\t};\t)$ and hence 
\begin{equation}
\bG(t;\t)=i\bG^{\rm R}(t;0)\bG(0;\t).
\label{newder9}
\end{equation}
Substituting Eq. (\ref{newder9}) into Eq. (\ref{newder4}) one gets 
\begin{equation}
\bG^{\lessgtr}(t;t')=\bG^{\rm R}(t;0)\bG^{\lessgtr}(0;0)
\bG^{\rm A}(0;t').
\label{newder11}
\end{equation}
Eq. (\ref{newder11}) coincides with Eq. (\ref{g<}), as it should.

\subsection{Interacting Systems in the Presence of an External 
Disturbance}
\label{intkel}

In the interacting case we keep track of the interactions by 
introducing a self-energy matrix. Then, Eq. (\ref{eomg}) becomes 
\begin{eqnarray}
\left\{i\frac{d}{dz}-\bcalE(z)-\bV(z)-
\bS^{\d}(z)\right\}\bG(z;z')
\nonumber \\ =\d(z-z')
+\int_{\g}d{\bar z}\;\bS_{\rm c}(z;{\bar z})\bG({\bar z};z'). 
\label{ieom1td}
\end{eqnarray}
Here $\bS^{\d}$ is the self-energy part which is 
local in time and it consists of a Hartree and an exchange 
term. The remaining part of the self energy $\bS_{\rm c}$ contains the contributions 
coming from the correlation and belongs to the Keldysh space:\cite{daniele} 
$$
\bS_{\rm c}(z;z')=\Theta(z,z')\bS^{>}(z;z')+\Theta(z',z)\bS^{<}(z;z').
$$
Like $\bG$, the self energy and its components are matrices in 
the one-electron labels. No simple expressions, like 
(\ref{gra})-(\ref{g<}), can now be directly obtained from the 
equation of motion and the Keldysh formalism is unavoidable.

A proper treatment of the initial correlations naturally leads to an 
extension of the Keldysh equations. The generalization was 
put forth by Wagner\cite{wagner} who obtained a minimal set of five 
independent integro-differential equations for the unknowns 
$\bG^{\rm R}$, $\bG^{\rm A}$, $\bG^{<}$ (or $\bG^{>}$) 
(or the Keldysh Green 
function $\bG^{\rm K}\equiv \bG^{>}+\bG^{<}$), 
$\bG^{\lceil}$ (or $\bG^{\rceil}$) 
and the thermal Green function $\bG$ with two imaginary arguments. 
In Appendix \ref{z} we exploit the results of the previous Subsection 
to prove that the solution for $\bG^{\lessgtr}$ can be written as 
\begin{equation}
\bG^{\lessgtr}(t;t')=\bG^{\rm R}(t;0)
\bG^{\lessgtr}(0;0)\bG^{\rm A}(0;t')+
\bDelta^{\lessgtr}(t;t'),
\label{GGGG} 
\end{equation}
where
\begin{eqnarray}
\bDelta^{\lessgtr}(t;t')&=&
 i\bG^{\rm R}(t;0)\bG^{>}(0;t')-
i\bG^{<}(t;0)\bG^{\rm A}(0;t')
\nonumber \\ &&-
\;\bG^{\rm R}(t;0)
\bG^{\rm K}(0;0)
\bG^{\rm A}(0;t')
\nonumber \\ && +
\left[\bG^{\rm R}\cdot\left[\bS^{\lessgtr}+
\bS^{\rceil}\star \bG\star \bS^{\lceil}
\right]\cdot \bG^{\rm A}\right](t;t').
\quad\;\;\,
\nonumber
\end{eqnarray}
This result clearly reduces to Eq. (\ref{newder11}) if the 
self energy vanishes since $\bG^{\rm R}(0;0)=
[\bG^{\rm A}(0;0)]^{\dag}=-i$.  
We observe that if the Green functions 
vanish when the 
separation of their time arguments goes to infinity, Eq. (\ref{GGGG})
yields a well known identity
\begin{equation}
\lim_{t,t'\ra\inf}\bG^{\lessgtr}(t;t')=
\left[\bG^{\rm R}\cdot\bS^{\lessgtr}\cdot \bG^{\rm A}\right](t;t').
\label{rsimp}
\end{equation}
Eq. (\ref{rsimp}) is well suited to study the \textit{ long-time} response of an 
interacting system subjected to an external time-dependent disturbance. 
On the other hand, if one is interested in the \textit{ short-time} 
response Eq. (\ref{GGGG}) cannot be simplified. In some cases it 
might be simpler to use an alternative approach. Below we 
propose an exact non-equilibrium Green function treatment based on TDDFT 
and discuss the relations to ordinary Mean Field approximations.

\subsection{Mean Field Theory and Relations to TDDFT}
\label{mftddft}

Any Mean Field Theory is a one-particle-like approximation in which 
each particle moves in an effective average potential independently of 
all other particles. The mean-field potential is local in time, 
meaning that $\bS_{\rm c}$ is discarded. Consequently, 
all the results of the Section \ref{nonint} can be reused provided we 
substitute $\bK$ 
by $\bK+\bS^{\d}$. Thus, 
no extra complications arise if we treat an interacting system at the 
Hartree-Fock level. To be specific, let us focus 
on the Coulomb interaction and on paramagnetic 
systems (so that the self energy and the Green function are diagonal 
in the spin indices). Then, it is natural to 
choose the one-electron index as the coordinate ${\bf r}$ of the 
particle and to split the self energy 
$\S^{\d}_{{\bf r},{\bf r}'}(z)\equiv \S^{\d}({\bf r},{\bf r}',z)$ as 
a sum of the Hartree and the exchange term 
$$
\S^{\d}({\bf r},{\bf r}',z)=V_{\rm H}({\bf r},z)\d({\bf r}-{\bf r}')+
\S_{\rm x}({\bf r},{\bf r}',z).
$$
For extended systems, the Hartree potential $V_{\rm H}$ and the Coulomb potential 
from the nuclei $V_{\rm n}$ are separately infinite but with a finite 
sum. Together with the external field $U$ these terms form the 
classical electrostatic potential $U_{\rm C}=U+V_{\rm H}+V_{\rm n}$.   
The Green function 
$G_{{\bf r},{\bf r}'}(z;z')\equiv G({\bf r},z;{\bf r}',z')$ 
can be obtained from the self-consistent solution of the 
equation of motion and the 
lesser/greater component can be written as 
\begin{eqnarray}
G^{\lessgtr}_{\rm MF}({\bf r},t;{\bf r}',t')&=&
\int d{\bar {\bf r}}d{\bar {\bf r}}'\;
G^{\rm R}_{\rm MF}({\bf r},t;{\bar {\bf r}},0)
\nonumber \\ &&\times
G^{\lessgtr}_{\rm MF}({\bar {\bf r}},0;{\bar {\bf r}}',0)
G^{\rm A}_{\rm MF}({\bar {\bf r}}',0;{\bf r}',t'),
\nonumber
\end{eqnarray}
where the subscript MF has been used to stress that it is a Mean Field 
approximate result. In the ordinary Many Body Theory 
one has to abandon the one-particle picture in order to improve 
the approximation beyond the Hartree-Fock level. This leads to 
a self energy nonlocal in time and hence to the complicated 
solution (\ref{GGGG}). 

In the case we only ask for the density $n({\bf r},t)=
-i G^{<}({\bf r},t;{\bf r},t)$ the original Density Functional 
Theory\cite{hk,ks} and its finite-temperature
generalization\cite{mermin}  has 
been extended to time-dependent phenomena.\cite{rg,litong} 
The theory applies only to 
those cases where the external disturbance is local in space, 
\textit{ i.e.}, $U_{{\bf r},{\bf r}'}(t)=\d({\bf r}-{\bf r}')U({\bf r},t)$. 
For $t>0$ we switch on an external 
potential $U({\bf r},t)$ to obtain a density $n({\bf r},t)$. The 
Runge-Gross theorem states that if we instead had switched on a 
different $U'({\bf r},t)$ [giving a different $n'({\bf r},t)$], then 
$n({\bf r},t)=n'({\bf r},t)$ implies $U({\bf r},t)=U'({\bf r},t)$. 
Thus $U({\bf r},t)$ is a unique functional of $n({\bf r},t)$. Runge 
and Gross also show that one can compute $n({\bf r},t)$ in a one-particle 
manner using an effective potential 
$$
U^{\rm eff}({\bf r},t)=U_{\rm C}({\bf r},t)+
v_{\rm xc}({\bf r},t).
$$
Here, $v_{\rm xc}$ accounts for exchange and correlations and is obtained
from an exchange-correlation action functional, $v_{\rm xc}({\bf r},t)=
\d A_{\rm xc}[n]/\d n({\bf r},t)$. In our earlier language this 
corresponds to an effective self energy which is local in 
both space and time. The TDDFT one-particle scheme 
corresponds to a fictitious Green function ${\cal G}({\bf r},z;{\bf 
r}',z')$ which satisfies the equations of motion (\ref{eomg}) 
with $[{\cal E}_{{\bf r},{\bf r}'}(z)+V_{{\bf r},{\bf r}'}(z)]$ replaced by 
$\d({\bf r}-{\bf r}')[-\nabla^{2}_{\bf r}/2+U^{\rm eff}({\bf r},z)]$. 
As a consequence we have
\begin{eqnarray}
{\cal G}^{\lessgtr}({\bf r},t;{\bf r}',t')&=&    
\int d{\bar {\bf r}}d{\bar {\bf r}}'\;
{\cal G}^{\rm R}({\bf r},t;{\bar {\bf r}},0)
\nonumber  \\ &&\times
{\cal G}^{\lessgtr}({\bar {\bf r}},0;{\bar {\bf r}}',0)
{\cal G}^{\rm A}({\bar {\bf r}}',0;{\bf r}',t').
\nonumber 
\end{eqnarray}
The fictitious $\bcalG$ will not in general give correct 
one-particle properties. However by definition $\bcalG^{<}$ gives 
the correct density
$$
n({\bf r},t)=-2i {\cal G}^{<}({\bf r},t;{\bf r},t) 
$$
(where the factor of 2 comes from spin).  
Also total currents are correctly given by TDDFT. If for instance 
$J_{\a}$ is the total current from a particular region $\a$ we have  
\begin{equation}
J_{\a}(t)=-e\int_{\a}d {\bf r}\;\frac{d}{dt}n({\bf r},t)
\label{cudft}
\end{equation}
where the space integral extends over the region $\a$ ($e$ 
is the electron charge).

The Density Functional Theory and the Runge-Gross extension refer specifically to the 
${\bf r}$ basis. However, the arguments remain valid if we instead 
consider the diagonal density $n_{i}=\bra c^{\dag}_{i}c_{i}\ket$ in 
some other basis provided the interactions commute with the diagonal 
density operator. The latter condition is essential for the 
Runge-Gross theorem. Thus, for instance, if the one-electron 
indices refer to a particular lead 
one can still use Eq. (\ref{cudft}) to calculate the corresponding total 
current (see Section \ref{interacting}).

For later references we now derive an expression for the 
lesser/greater Green function in the linear approximation. We consider 
the partition-free system described in the one-particle scheme of Mean 
Field Theory or TDDFT. Let $\d \bU^{\rm eff}(t)$ be the 
small time-dependent effective perturbation and 
$\d \bG^{\rm R,A}=\bG_{0}^{\rm R,A}\cdot 
\d\bU^{\rm eff}\cdot 
\bG_{0}^{\rm R,A}$ be the first 
order variation of the retarded and advanced Green functions with 
respect to their equilibrium counterparts $\bG_{0}^{\rm R,A}$. Then, from 
Eq. (\ref{newder11}) we get 
\begin{eqnarray}
\d \bG^{\lessgtr}(t;t')
\quad\quad\quad\quad\quad
\quad\quad\quad\quad\quad
\quad\quad\quad\quad\quad
\quad\quad\quad
\label{gamdp2} \\
=\int d{\bar t}\;
\bG_{0}^{\rm R}(t;{\bar t})
\d \bU^{\rm eff}({\bar t})\bG^{\lessgtr}(0;0)\bG_{0}^{\rm R}({\bar t};0)
\bG_{0}^{\rm A}(0;t')
\nonumber \\ \;\; +
\int d{\bar t}\;\bG_{0}^{\rm R}(t;0)
\bG_{0}^{\rm A}(0;{\bar t})
\bG^{\lessgtr}(0;0)\d \bU^{\rm eff}({\bar t})\bG_{0}^{\rm A}({\bar t};t'),
\nonumber
\end{eqnarray}
where we have taken into account that $\bG_{0}^{\rm R,A}$ 
commutes with $\bG^{\lessgtr}(0;0)$. 
The above expression takes an elegant form when $t'=t$. Indeed, for any 
$t>{\bar t}>0$ one has $\bG_{0}^{\rm R}({\bar t};0)
\bG_{0}^{\rm A}(0;t)=-i\bG_{0}^{\rm A}({\bar t};t)$ and 
$\bG_{0}^{\rm R}(t;0)\bG_{0}^{\rm A}(0;{\bar t})=
i\bG_{0}^{\rm R}(t;{\bar t})$.
Since the integrands in Eq. (\ref{gamdp2}) vanish for ${\bar t}>t$ due 
to the $\Theta$ function in $\bG_{0}^{\rm R}$ in the first term and in 
$\bG_{0}^{\rm A}$ in the second term, we conclude that for any positive 
time $t$ 
\begin{equation}
\d \bG^{\lessgtr}(t;t)=-i\int d{\bar t}\; 
\bG_{0}^{\rm R}(t;{\bar t})[\d \bU^{\rm eff}({\bar t}),\bG^{\lessgtr}(0;0)]
\bG_{0}^{\rm A}({\bar t};t).
\label{ga20df;}
\end{equation}
We shall use this equation later on to calculate the linear current 
response in noninteracting and interacting 
resonant tunneling systems.

\section{Noninteracting Resonant Tunneling Systems}
\label{model}

As a first application of the partition-free approach we  
study the time-dependent current response of a noninteracting 
resonant tunneling system. For the sake of simplicity the 
central device will be modeled by a single localized level.
All the results of this 
Section can be generalized to the case of a multi-level noninteracting 
central device without any conceptual complications. There are many 
different geometries one can conceive beyond a one-level model, 
\textit{ e.g.}, a double quantum dots model,\cite{ziegler}  a quantum wire 
coupled to a quantum dot,\cite{liang} a one-dimensional quantum-dot 
array\cite{shangguam} or a mesoscopic multi-terminal system.\cite{sun} 
However, the present paper is not intended to give a description of a 
series of applications. Rather, we 
prefer to illustrate how the partition-free approach works in a  
simple noninteracting model. We also emphasize that \textit{all} the 
results of this Section remain valid in the interacting case if the 
bare external potential is replaced by the \textit{exact}  
effective potential of TDDFT, see Section \ref{interacting}. 

The whole system is described by a quadratic Hamiltonian 
\begin{eqnarray}
H_{0}&=&\sum_{k\a}\ve_{k\a}c^{\dag}_{k\a}c_{k\a}+\ve_{0}c^{\dag}_{0}c_{0}
\label{qham}  \\ &&+
\sum_{k\a}V_{k\a}[c^{\dag}_{k\a}c_{0}+c^{\dag}_{0}c_{k\a}]
\equiv
\sum_{mn}T_{m,n}c^{\dag}_{m}c_{n},
\nonumber 
\end{eqnarray}
where $\a=L,R$ denotes the left, right lead and $m,n$ are collective 
indices for $k\a$ and 0. We assume the system in thermodynamic 
equilibrium at a given inverse temperature $\b$ and chemical 
potential $\m$ before the time-dependent perturbation 
\begin{eqnarray}
H_{U}(t)&=&\sum_{k\a}U_{k\a}(t)c^{\dag}_{k\a}c_{k\a}+
U_{0}(t)c^{\dag}_{0}c_{0}
\nonumber \\ &\equiv&
\sum_{mn}U_{m,n}(t)c^{\dag}_{m}c_{n}
\nonumber
\end{eqnarray}
is switched on. 
In principle the time-dependent perturbation may have 
off-diagonal matrix elements. In order to model a uniform 
potential deep inside the electrodes such off-diagonal terms must be 
of lower order with respect to the system size. However their 
inclusion is trivial and it does not lead to any qualitative changes. 

The current from the $\a$ contact through the $\a$ barrier to the 
central region can be calculated from the time evolution of the 
occupation number operator $N_{\a}$ of the $\a$ contact. 
From the obvious generalization of Eq. (\ref{cudft}) 
one readily finds
\begin{eqnarray}
J_{\a}(t)&=&
2e\sum_{k}\;\Re\left[G^{<}_{0,k\a}(t;t)\right]V_{k\a}
\label{current1} \\ &=&
2e\sum_{k}\;\Re\left[\bG^{\rm R}(t;0)\bG^{<}(0;0)
\bG^{\rm A}(0;t)\right]_{0,k\a}V_{k\a}.
\nonumber
\end{eqnarray}
The above expression is manifestly gauge-invariant. Indeed, if 
$U_{m,n}(t)\ra U_{m,n}(t)+\d_{m,n}\chi(t)$ then 
$\bG^{\rm R}(t;0)\ra 
{\rm e}^{-i\int_{0}^{t}\chi(\bar{t})d\bar{t}}\bG^{\rm R}(t;0)$ while 
$\bG^{\rm A}(0;t)\ra 
{\rm e}^{i\int_{0}^{t}\chi(\bar{t})d\bar{t}}\bG^{\rm A}(0;t)$ 
and the time-dependent shift $\chi(t)$ 
has no effect on the current response. In the same way it is invariant 
under a simultaneous shift of $\m$ and the initial 
potential.

The matrix $\bG^{<}(0;0)$ can be written as\cite{lw} 
$$
\bG^{<}(0;0)=\int_{\G}\frac{d\z}{2\p}
\frac{f(\z){\rm e}^{\eta\z}}{\z-\bcalE-\bV},
$$
where $\G$ is the contour surrounding all the Matzubara frequencies 
$\w_{n}=(2n+1)\p i/\b+\m$ clockwise [see Fig. \ref{lwcontour} in 
Appendix \ref{a2}] while $\eta$ is an infinitesimally small positive 
constant. 
It is therefore convenient to define the kernel  
\begin{equation}
Q_{\a}(\z;t)=\sum_{k}\left[
\bG^{\rm R}(t;0)\bG(\z)\bG^{\rm A}(0;t)\right]_{0,k\a}V_{k\a},
\label{qalfa}
\end{equation} 
with $\bG(\z)=[\z-\bcalE-\bV]^{-1}$, 
and to write the current in the form
\begin{equation}
J_{\a}(t)=
2e\;\Re\left[\int_{\G}\frac{d\z}{2\p}f(\z){\rm e}^{\eta\z}
Q_{\a}(\z;t)\right].
\label{current2}
\end{equation}

It is worth noticing that the partitioned approach  
leads to Eq. (\ref{current1}) with ${\bf g}^{<}(0;0)=if(\bcalE)$ in place of 
$\bG^{<}(0;0)=if(\bcalE+\bV)$. It is our intention to clarify under what 
circumstances, if any, the long-time behavior of the time-dependent 
current is not affected by this replacement. 

As a side remark we also observe that $J_{\a}(t=0)$ in 
Eq. (\ref{current1}) correctly vanishes. 
Letting $|\l_{n}\ket$ and $\l_{n}$ be the eigenvectors and eigenvalues  
of $\bT=\bcalE+\bV$,  we have
$$
J_{\a}(0)=
-2e\sum_{k}\sum_{n}\Im\left[\bra 0|\l_{n}\ket f(\l_{n})
\bra\l_{n}|k\a\ket\right]V_{k\a}=0,
$$
since $\bra 0|\l_{n}\ket$ and $\bra\l_{n}|k\a\ket$ can always be 
chosen as real quantities for systems with time reversal symmetry. 

\subsection{Step-Like Modulation}

The first exactly solvable model we wish to consider is a step-like 
modulation, \textit{ i.e.}, $U_{m,n}(t)=\Theta(t)U_{m,n}$. From Eq. (\ref{gra}) 
it follows that for any $t>0$
$$
\bG^{\rm R}(t;0)=-i{\rm e}^{-i[\bcalE+\bU+\bV]t}
\equiv \int\frac{d\w}{2\p}{\rm e}^{-i\w t}\bG^{\rm R}(\w), 
\label{grt>0}
$$
and $\bG^{\rm A}(0;t)=[\bG^{\rm R}(t;0)]^{\dag}$. 
The device component of $\bG^{\rm R,A}(\w)$ can be written as 
\begin{equation}
G^{\rm R,A}_{0,0}(\w)=\frac{1}{\w-\tilde{\ve}_{0}-\S^{\rm R,A}(\w)\pm i\eta},
\label{g00}
\end{equation}
where 
$\tilde{\ve}_{0}=\ve_{0}+U_{0}$. Here, $\S^{\rm R,A}(\w)=
\sum_{\a}\S^{\rm R,A}_{\a}(\w)$ is the retarded/advanced self energy induced by 
back and forth virtual hopping processes from the localized level to 
the leads and is given by
\begin{equation}
\S^{\rm R,A}_{\a}(\w)=
\sum_{k}\frac{V_{k\a}^{2}}{\w-\tilde{\ve}_{k\a}\pm i\eta},
\label{ser}
\end{equation}
where we have used the short-hand notation $\tilde{\ve}_{k\a}=
\ve_{k\a}+U_{k\a}$. 
%

\subsubsection{$U_{k\a}=U_{\a}$: Steady-State Current}

If $U_{k\a}=U_{\a}$ the energy levels of the $\a$ lead are 
equally shifted. 
From Eq. (\ref{current1}) it follows that we need to estimate the 
matrix elements 
$G^{\rm R}_{0,0}(t;0)$, $G^{\rm R}_{0,k\a}(t;0)$ of the retarded 
Green function and the two contractions  
$\sum_{k}G^{\rm A}_{0,k\a}(0;t)V_{k\a}$,  
$\sum_{k}G^{\rm A}_{k'\a',k\a}(0;t)V_{k\a}$ in the long-time limit. 
We assume that  $\S_{\a}^{\rm R,A}(\w)$ is a smooth 
function for all real $\w$. Then, using the the Riemann-Lebesgue theorem 
one can prove (see Appendix \ref{a1}) that the kernel $Q_{\a}(\z;t)$ 
has the following asymptotic behavior
\begin{eqnarray}
\lim_{t\ra\inf}Q_{\a}(\z;t)=
\int\frac{d\ve}{2\p}\frac{\G_{\a}(\ve)}{\z-\ve+ U_{\a}}
G^{\rm R}_{0,0}(\ve)\quad\quad\quad\quad
\label{INTERM} \\ +
\sum_{\a'}\int\frac{d\ve}{2\p}
\frac{\G_{\a'}(\ve)}{\z-\ve+ U_{\a'}}
\left|G^{\rm R}_{0,0}(\ve)\right|^{2}
\S^{\rm A}_{\a}(\ve),
\nonumber
\end{eqnarray}
where
\begin{equation}
\G_{\a}(\ve)=-2\Im[\S^{\rm R}_{\a}(\ve)]=
2\p\sum_{k}\d(\ve-\tilde{\ve}_{k\a})V^{2}_{k\a}.
\label{gammalfa}
\end{equation}
In Eq. (\ref{INTERM}) the r.h.s. has a simple pole structure in the 
$\z$ variable and therefore the integration 
along the $\G$ contour can be easily performed. 
Using the identity 
$\int_{\G}(d\z/2\p)f(\z){\rm e}^{\eta\z}
(\z-\ve)^{-1}=if(\ve)$
the stationary current $J^{(\rm S)}_{\a}\equiv
\lim_{t\ra\inf}J_{\a}(t)$ has the following expression 
\begin{eqnarray}
J^{(\rm S)}_{R}&=&-e\int \frac{d\ve}{2\p}
\frac{\G_{R}(\ve)\G_{L}(\ve)}{[\ve-\tilde{\ve}_{0}-\L(\ve)]^{2}+
[\G(\ve)/2]^{2}} \nonumber \\ 
&&\quad\quad \times
[f(\ve-U_{L})-f(\ve-U_{R})]
=-J^{(\rm S)}_{L},
\label{js0}
\end{eqnarray}
where $\L(\ve)=\Re[\S^{\rm R}(\ve)]$ is the Hilbert transform 
of $\G(\ve)=\sum_{\a}\G_{\a}(\ve)$:
\begin{equation}
\L(\w)=P\int\frac{d\w'}{2\p}\frac{\G(\w')}{\w-\w'}.
\label{lambdaalfa}
\end{equation}
It is of interest to note that the dependence on the bias $U_{\a}$ appears 
not only in the distribution function $f$ but also in the 
quantities $\G$ and $\L$, see Eqs. (\ref{gammalfa})-(\ref{lambdaalfa}). 
The dependence of the self energy on the level-shifts is physical since
when the particle 
visits the reservoirs experiences the applied potential. We also 
remark that Eq. (\ref{js0}) is of the Landauer type.\cite{landauer}  
More generally the Landauer formula is valid for any mesoscopic 
device provided it is noninteracting. 
This result agrees with the one obtained in the partitioned 
approach by Jauho and coworkers.\cite{jauho,haug} There 
the leads are decoupled from the 
central device and in thermal equilibrium at different chemical 
potential $\m_{L}$ and $\m_{R}$ and inverse temperature $\b_{L}$ and 
$\b_{R}$ in the remote past. In order to preserve charge 
neutrality each energy level $\ve_{k\a}$ must be shifted by $\m_{\a}-\m$ 
where $\m$ is the chemical potential of the two undisturbed leads. 
The stationary 
current is then obtained by switching on the contacts, \textit{ i.e.}, the 
hybridization part of the Hamiltonian. By 
tuning $\b_{R}=\b_{L}=\b$ and $\m_{R}-\m_{L}=U_{R}-U_{L}$ the current 
is given by Eq. (\ref{js0}).

To summarize we have found that for 
noninteracting leads a steady state develops in the long-time limit 
whenever 1) The one-body levels of the charge reservoirs form a 
continuum and 2) The self energy due to the hopping term is a smooth 
function. Under these hypotheses the time-translational invariance is 
restored by means of a \textit{dephasing mechanism}. 
The comparison of our result with the one obtained in the partitioned 
scheme provides the criteria of equivalence: besides the 
tuning $\m_{L}-\m_{R}=U_{L}-U_{R}$ one needs to shift the levels of 
the $\a$ reservoir by $\m_{\a}-\m$.

\subsubsection{$U_{k\a}=U_{\a}$: Time-Dependent Current in the Wide-Band Limit}

The calculation of the stationary current is greatly simplified by the 
long-time behavior of the various terms coming from  
Eq. (\ref{current1}). However, as far as we are 
interested in the current at any finite time we need to specify the 
structure of the retarded (advanced) self energy. Here, we consider 
the so called wide-band limit where the level-width functions 
$\G_{\a}(\w)\equiv 2\g_{\a}$ are assumed to be constant and hence, 
from Eq. (\ref{lambdaalfa}), $\L_{\a}(\w)=0$. In this case 
$G^{\rm R}_{0,0}(\w)$ has a simple-pole structure and the calculations 
are slightly simplified. We emphasize that what follows is the first explicit 
result of a time-dependent current in a model system in the 
framework of a partition-free approach and  
therefore also a simple model could be of some interest. 
Without loss of generality we can always choose $\ve_{0}=0$; for the sake of 
simplicity we also consider $U_{0}=0$. 
We defer the reader to Appendix \ref{a2} for the details. Here, we just 
write down the final result for $J_{\a}(t)$:
\begin{eqnarray}
&&J_{\a}(t)=J_{\a}^{(\rm S)}-4e\g_{\a}{\rm e}^{-\g t}
\int\frac{d\w}{2\p}f(\w)
\label{DJFNFCO} \\ 
&&\times\left[U_{\a}
\;\Im\left\{\frac{{\rm e}^{i(\w+U_{\a})t}}
{(\w+i\g)(\w+U_{\a}+i\g)}\right\}+
\sum_{\a'}\g_{\a'}U_{\a'}\;
\quad
\right.\nonumber \\ &&\left.\times
\frac{U_{\a'}{\rm e}^{-\g t}+2\w\cos[(\w+U_{\a'})t]+2\g\sin[(\w+U_{\a'})t]}
{[\w^{2}+\g^{2}][(\w+U_{\a'})^{2}+\g^{2}]}\right],
\nonumber 
\end{eqnarray}
where $J_{\a}^{(\rm S)}$ is the stationary current of Eq. (\ref{js0}) 
and $\g=\g_{R}+\g_{L}$. 
One can easily check that 
1) For $t\ra\inf$ Eq. (\ref{DJFNFCO}) yields the result in Eq. (\ref{js0}), 2) 
For $t=0$ the current vanishes, that is $J_{\a}(0)=0$ and 3) For $U_{L}=U_{R}=0$ 
the current vanishes for any $t$. 
Eq. (\ref{DJFNFCO}) can be rewritten in a more physical and compact way if 
we exploit the particle-number conservation. 
Denoting by $n_{0}$ the particle number operator in the central device we 
have
$$
J_{R}(t)+J_{L}(t)=e\frac{d}{dt}\bra n_{0}\ket,
$$
so that
\begin{eqnarray}
J_{R}(t)=J_{R}^{(\rm S)}+e\frac{\g_{R}}{\g}\frac{d}{dt}\bra n_{0}\ket-
4e\frac{\g_{R}\g_{L}}{\g}{\rm e}^{-\g t}
\int\frac{d\w}{2\p}
\quad
\nonumber \\\times
\Im\left\{\frac{f(\w)}{\w+i\g}\left[
U_{R}\frac{{\rm e}^{i(\w+U_{R})t}}{\w+U_{R}+i\g}-
U_{L}\frac{{\rm e}^{i(\w+U_{L})t}}{\w+U_{L}+i\g}
\right]\right\};
\nonumber
\end{eqnarray}
$J_{L}(t)$ is obtained by exchanging $R\leftrightarrow L$ in the 
r.h.s. of the above expression. Therefore, 
$J_{R}(t)\neq -J_{L}(t)$ for any finite 
time $t$, even in the symmetric case $\g_{R}=\g_{L}$; 
the time derivative of $\bra n_{0}\ket$ contributes to 
$J_{R}$ and $J_{L}$ in the same way. 

Our formula for the nonlinear transient current clearly differ 
from the one obtained by Jauho \textit{et al.}\cite{jauho} in the partitioned 
scheme. Indeed, the prescribed integration along the imaginary axis 
gives extra terms (see Appendix \ref{a2}) which are absent if the system 
is uncontacted for negative times. 
We have explicitly verified that by discarding these  
terms our formula reduces to the one obtained in the partitioned 
scheme. For long times, the extra terms vanish and our scheme 
reproduces the earlier steady-state results.

If one of the two leads does not undergo any level shift, 
\textit{e.g.} $U_{R}=0$, from Eq. (\ref{DJFNFCO}) we get
\begin{eqnarray}
J_{R}(t)=J_{R}^{(\rm S)}
-4e\g_{R}\g_{L}U_{L}{\rm e}^{-\g t}\int\frac{d\w}{2\p}f(\w)
\quad\quad\quad\quad
\label{tdcr} \\ \times
\frac{U_{L}{\rm e}^{-\g t}+2\w\cos[(\w+U_{L})t]+2\g\sin[(\w+U_{L})t]}
{[\w^{2}+\g^{2}][(\w+U_{L})^{2}+\g^{2}]}.
\nonumber
\end{eqnarray}

\begin{figure}[htbp]
\begin{center}
\includegraphics*[scale=1.0]{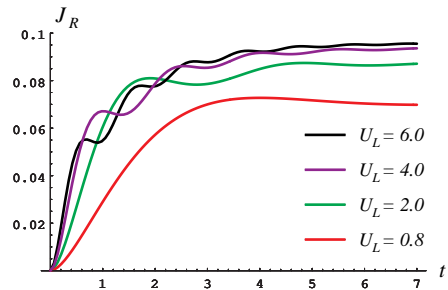}
\caption{{\footnotesize Time-dependent current $J_{R}(t)$ for 
    different values of the applied bias $U_{L}=0.8,\;2.0,\;4.0$ 
    and $6.0$. The numerical integration has been done with 
    $\g_{R}=\g_{L}=0.2$, $\m=0$ and zero temperature.}}
\label{transiente1}
\end{center}
\end{figure}

The transient behavior of the time-dependent quantity $J_{\a}(t)-
J_{\a}^{(\rm S)}$ is not simply an exponential decay.
In Fig. \ref{transiente1} we have plotted $J_{R}(t)$ in Eq. (\ref{tdcr}) 
versus $t$ for 
different values of the applied bias $U_{L}$ at zero temperature. 
The current strongly depends on $U_{L}$ for small $U_{L}$ 
while it is fairly independent of it in the strong bias regime; using 
the parameter specified in the caption, the time-dependent current has 
essentially the same shape for any $U_{L}\gtrsim 8$. 

\begin{figure}[htbp]
\begin{center}
\includegraphics*[scale=1.0]{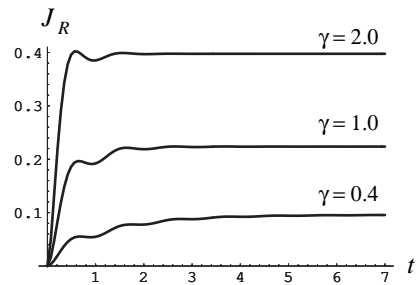}
\caption{{\footnotesize Time-dependent current for $U_{L}=6$, $\m=0$ and 
    zero temperature versus time for three different values of the 
    line widths $\g_{R}=\g_{L}=0.2,\;0.5$ and $1.0$.}}
\label{transiente2}
\end{center}
\end{figure}

In Fig. \ref{transiente2} the current $J_{R}(t)$ is plotted for 
different values of the total line width $\g$ and for a fixed value 
$U_{L}=6$ of the applied bias. As expected, the larger is $\g$ the 
bigger is the slope of the current in $t=0$. 

\begin{figure}[htbp]
\begin{center}
\includegraphics[scale=0.7]{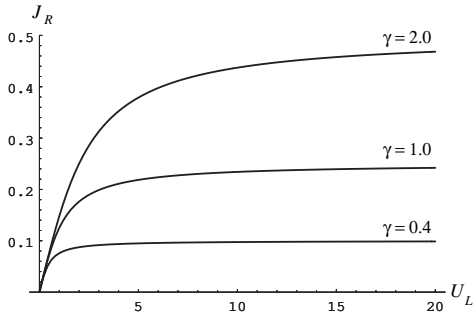}
\caption{{\footnotesize Stationary current versus the applied bias at 
zero temperature and chemical potential 
    for three different values of the 
    line widths $\g_{R}=\g_{L}=0.2,\;0.5$ and $1.0$.}}
\label{stationary}
\end{center}
\end{figure}

Finally, in Fig. \ref{stationary} we report the trend of the 
stationary current $J_{R}^{(\rm S)}$ as a function of the bias 
$U_{L}$ for three different choices of the level widths. As one can 
see the bigger is $\g$ and the wider is the range of validity of the 
Ohm law.

\subsection{Arbitrary Modulation: Theorem of Equivalence}
\label{toe}

We have shown that the steady-state current induced by a step-like 
modulation does not change 
if one uses the $\bG^{<}$ of the partitioned approach, 
given by the first term on 
the r.h.s. of Eq. (\ref{newder4}), in 
place of the one coming from the partition-free approach. 
This reasonable result is now \textit{ proved} 
and not simply \textit{ postulated}. The equivalence between the two 
expressions for the current is of special importance 
since it is much easier to work in the partitioned scheme. 
However, it has been proved only for step-like 
modulations with $U_{k\a}=U_{\a}$. Here, we prove that the above 
equivalence remains true under very general assumptions. To this end 
we consider the quantity
\begin{equation}
\S^{\rm R,A}_{\a,{\cal V}}(t;t')=
\sum_{k}\,\tg^{\rm R,A}_{k\a}(t;t')
{\cal V}^{2}_{k\a}
\label{fse}
\end{equation}
where ${\cal V}$ is an arbitrary complex function of $k\a$.  
Then, the following theorem holds 

\textit{ \underline{Theorem of Equivalence}: If 
\begin{equation}
\lim_{t\ra\inf}\S^{\rm R}_{\a,{\cal V}}(t;t')=
\lim_{t\ra\inf}\S^{\rm A}_{\a,{\cal V}}(t';t)=0
\label{limco}
\end{equation}
for any nonsingular ${\cal V}$, then
\begin{equation}
\lim_{t\ra\inf}
[Q_{\a}(\z;t)-q_{\a}(\z;t)]=0,
\label{eoe}
\end{equation}
where $q_{\a}(\z;t)\equiv\sum_{k}
[\bG^{\rm R}(t;0){\bf g}(\z)\bG^{\rm A}(0;t)]_{0,k\a}V_{k\a}$, 
and ${\bf g}(\z)=[\z-\bcalE]^{-1}$ is the uncontacted Green function.}

Eq. (\ref{eoe}) says that if we apply the same time-dependent 
perturbation the same asymptotic current 
will emerge in the partitioned and partition-free approaches. 

\textit{ \underline{Proof}}: In terms of the self energy 
$\S^{\rm R}_{V}=\sum_{\a}\S^{\rm R}_{\a,V}$, the equation of motion 
for $G^{\rm R}_{0,0}$ takes the form
$$
\left\{i\frac{d}{dt}-\ve_{0}(t)\right\}
G^{\rm R}_{0,0}(t;t')-
[\S^{\rm R}_{V}\cdot G^{\rm R}_{0,0}](t;t')=\d(t-t'),
$$
where the symbol $\cdot$ denotes the real-time convolution. 
We now consider the limit $t\ra\inf$. The hypothesis 
(\ref{limco}) implies 
\begin{equation}
\lim_{t\ra\inf}G^{\rm R}_{0,0}(t;t')=0,
\label{gr00t0}
\end{equation} 
which in turn implies that 
$[\S^{\rm R}_{V}\cdot G^{\rm R}_{0,0}](t;t')
\stackrel{t\ra\inf}{\ra}0$.
Furthermore, from the 
Dyson equation for $G^{\rm A}_{0,k\a}$ we find
\begin{equation}
\sum_{k}G^{\rm A}_{0,k\a}(t';t)V_{k\a}=
[G^{\rm A}_{0,0}\cdot\S^{\rm A}_{\a,V}](t';t)
\stackrel{t\ra\inf}{\ra}0.
\label{asycon}
\end{equation}
We note that the above two asymptotic relations have been obtained for 
$t'=0$ in the special case of a step-like modulation, see 
Eq. (\ref{asymg00}). Here, we have shown that they 
hold in a more general context. As a consequence of these two 
identities, the asymptotic difference $[Q_{\a}(\z;t)-q_{\a}(\z;t)]$ 
can be written as
$$
\left[G^{\rm R}_{0,0}\cdot\S^{\rm R}_{\cal V}
\right](t;0)\times G_{0,0}(\z)\times\S^{\rm A}_{\cal V}\cdot
\left[\d+G^{\rm A}_{0,0}\cdot\S^{\rm A}_{\a,V}\right](0;t).
$$
Here, $\S^{\rm R}_{\cal V}=\sum_{\a}\S^{\rm R}_{\a,\cal V}$ is given by 
Eq. (\ref{fse}) with ${\cal V}^{2}_{k\a}=V^{2}_{k\a}/(\z-\ve_{k\a})$.
Since $\z\in\G$, $\Im[\z]\neq 0$ and hence ${\cal V}^{2}_{k\a}$ is 
nonsingular, meaning that Eq. (\ref{limco}) holds. Eq. (\ref{limco}) 
together with Eq. (\ref{asycon}) imply the equation of 
equivalence (\ref{eoe}).

As a simple application of the Theorem of Equivalence one can 
calculate the stationary current for an arbitrary step-like modulation. 
The quantity $q_{\a}(\z;t)$ is simply given by the first two terms of 
Eq. (\ref{statcurint}). Both have a simple pole structure in the $\z$ 
variable and we can perform the integration along the contour $\G$. 
Using the definition in Eq. (\ref{gammalfa}), with 
$\tilde{\ve}_{k\a}=\ve_{k\a}+U_{\a}(\ve_{k\a})$, one 
obtains
\begin{eqnarray}
J^{(\rm S)}_{R}&=&-e\int\frac{d\ve}{2\p}f(\ve)
\label{stacg} \\ 
&&\times\left\{\G^{(0)}_{L}(\ve)\G_{R}(\ve+U_{L}(\ve))
\left|G^{\rm R}_{0,0}(\ve+U_{L}(\ve))
\right|^{2}\right.
\nonumber \\ &&\left.+
\G^{(0)}_{R}(\ve)\G_{L}(\ve+U_{R}(\ve))
\left|G^{\rm R}_{0,0}(\ve+U_{R}(\ve))
\right|^{2}
\right\}
\nonumber
\end{eqnarray}
The quantity 
$\G^{(0)}_{\a}(\ve)\equiv 2\p\sum_{k}V_{k\a}^{2}\d(\ve-\ve_{k\a})$ is 
the equilibrium line width. Eq. (\ref{stacg}) reduces to Eq. (\ref{js0}) 
if $U_{\a}(\ve)=U_{\a}$ since in this case 
$\G^{(0)}_{\a}(\ve-U_{\a})=\G_{\a}(\ve)$.

In the noninteracting case it is reasonable to assume that Eq. (\ref{stacg}) 
yields the steady-state current even for an arbitrary time-dependent 
disturbance such that $\lim_{t\ra\inf}U_{k\a}(t)=U_{k\a}$ and 
$\lim_{t\ra\inf}U_{0}(t)=U_{0}$. In the next Section we shall prove 
that the asymptotic current has no memory and depends only on the 
asymptotic value of the external perturbation.

\subsection{Memory Loss Theorem}

If the condition (\ref{limco}) of the Theorem of Equivalence is 
fulfilled, the asymptotic value of the nonlinear time-dependent current in 
Eq. (\ref{current2}) simplifies to
\begin{eqnarray}
J_{\a}(t)=&&
2e\;\Re\left[\int_{\G}\frac{d\z}{2\p}f(\z){\rm e}^{\eta\z}
q_{\a}(\z;t)\right]\label{asymcur} \\ 
&&=2e\;\Re\left[
\sum_{k\b}G^{\rm R}_{0,k\b}(t;0)\tg^{<}_{k\b}(0;0)
 \right.
\nonumber \\ &&
\quad\quad\quad\quad
\quad\quad\left.\times
\sum_{k'}G^{\rm A}_{k\b,k'\a}(0;t)V_{k'\a}\right].
\nonumber
\end{eqnarray}

We note in passing that expressing $G^{\rm R}_{0,k\b}$ and 
$G^{\rm A}_{k\b,k'\a}$ in terms of $G^{\rm R}_{0,0}$ and 
$G^{\rm A}_{0,0}$ respectively, Eq. (\ref{asymcur}) can be rewritten 
in terms of $\S^{<}_{\a,V}=\sum_{k}\tg^{<}_{k\a}V^{2}_{k\a}$
$$
J_{\a}(t)=2e\;\Re\left\{
[G^{\rm R}_{0,0}\cdot\S^{<}_{\a,V}](t;t)+
[G^{<}_{0,0}\cdot\S^{\rm A}_{\a,V}](t;t)
\right\},
$$
where the asymptotic relation $G^{<}_{0,0}=\sum_{\a}G^{\rm 
R}_{0,0}\cdot \S^{<}_{\a,V}\cdot G^{\rm A}_{0,0}$ has been used
[see Eq. (\ref{rsimp})]. This 
agrees with the result obtained by Wingreen \textit{et 
al.}\cite{wingreen,jauho} in the partitioned approach, as it should.

In general $J_{\a}(t\ra\inf)$ is not a constant unless the external 
perturbation tends to a constant in the distant future. In this case 
the following theorem holds

\textit{ \underline{Memory Loss Theorem}: If 
$$
\lim_{t\ra\inf}U_{\a}(\ve,t)=U_{\a}(\ve),
\quad
\lim_{t\ra\inf}U_{0}(t)=U_{0}
$$
the current $J_{\a}(t)$ tends to a constant, given by 
Eq. (\ref{stacg}), in the long-time limit.}

\textit{ \underline{Proof}}: Is convenient to denote with ${\bar G}$ and 
$\bar{\tg}$ the Green  
functions corresponding to the step-like modulation with 
coefficients $U_{\a}(\ve)$ and $U_{0}$. We have already shown 
that in the long-time limit 
Eq. (\ref{asymcur}) yields Eq. (\ref{stacg}) if $G^{\rm 
R,A}=\bar{G}^{\rm R,A}$. The Memory Loss 
Theorem is then proved if 
\begin{equation}
\lim_{t\ra\inf}\frac{\bar{G}^{\rm R}_{0,k\b}(t;0)}
{G^{\rm R}_{0,k\b}(t;0)}=
{\rm e}^{i\D_{k\b}}=
\lim_{t\ra\inf}\frac{\sum_{k'}G^{\rm A}_{k\b,k'\a}(0;t)V_{k'\a}}
{\sum_{k'}\bar{G}^{\rm A}_{k\b,k'\a}(0;t)V_{k'\a}}
\label{wie3of3}
\end{equation}
for some real constant $\D_{k\b}$ . 

According to Eq. (\ref{gr00t0}), the device component of the 
retarded Green function $G^{\rm R}_{0,0}(t\ra\inf;t')$ vanishes for any 
finite $t'$. Since $\lim_{t,t'\ra\inf}
[\tg_{0}^{\rm R}(t;t')/\bar{\tg}_{0}^{\rm R}(t;t')]=1$, from  
$G^{\rm R}_{0,0}=\tg_{0}^{\rm R}+\tg_{0}^{\rm R}\cdot\S^{\rm 
R}_{V}\cdot G^{\rm R}_{0,0}$  it follows that
$\lim_{t,t'\ra\inf}[G^{\rm R}_{0,0}(t;t')/\bar{G}^{\rm R}_{0,0}(t;t')]=
1$. 
Let us now consider the Dyson equation
\begin{equation}
G^{\rm R}_{0,k\a}(t;t')=\int d\bar{t}\;
G^{\rm R}_{0,0}(t;\bar{t})V_{k\a}
\tg^{\rm R}_{k\a}(\bar{t};t'),
\label{gr0kacp}
\end{equation}
with $t'=0$ and $t\ra\inf$. Since the integrand vanishes for any finite 
$\bar{t}$, 
we can substitute $G^{\rm R}_{0,0}$ with $\bar{G}^{\rm R}_{0,0}$. 
Furthermore, since the applied bias tends to a constant in the distant 
future, $\lim_{t\ra\inf}[\tg^{\rm R}_{k\a}(t;0)/\bar{\tg}^{\rm R}_{k\a}(t;0)]=
{\rm e}^{-i\D_{k\a}}$ for some real quantity $\D_{k\a}$. The l.h.s. of
Eq. (\ref{wie3of3}) is then proved. A similar reasoning leads to the r.h.s. 
of Eq. (\ref{wie3of3}).

\subsection{Linear Response in the Wide-Band Limit}

In the case of small time-dependent perturbations,  
one can use Eq. (\ref{ga20df;}) to calculate the lesser Green 
function. In order to carry on the calculations 
analytically we consider the wide-band limit and we choose 
$\d U_{k\a}(t)=\d U_{\a}(t)$. For 
simplicity we omit the subscript 0 in the retarded 
and advanced equilibrium Green functions. By explicitly 
writing down the matrix product in Eq. (\ref{ga20df;}) one readily 
realize that we have to calculate the functions 
$G_{0,0}^{\rm R}(t-{\bar t})$, 
$G_{0,k'\a'}^{\rm R}(t-{\bar t})$, 
$\sum_{k}V_{k\a}G_{0,k\a}^{\rm A}({\bar t}-t)$ and 
$\sum_{k}V_{k\a}G_{k'\a',k\a}^{\rm A}({\bar t}-t)$. They are 
easily obtained from Eqs. (\ref{az,`/}) 
by simply replacing $\tilde{\ve}\ra\ve$ and $t\ra t-{\bar t}$. 
The calculations are rather similar to those already performed 
to derive the expression (\ref{DJFNFCO}) and they are left to the reader. 
Denoting by $\d J_{\a}(t)$ the time-dependent current in the linear 
regime one ends up with 
\begin{eqnarray}
\d J_{\a}(t)=4e\g_{\a}\Re\left\{
\int_{0}^{t}d{\bar t} \;
\int\frac{d\w}{2\p}f(\w)
\frac{{\rm e}^{i(\w-\W_{0})(t-{\bar t})}}{\w-\W_{0}}\right.
\nonumber\\ \left.
\times\left[\d U_{0}({\bar t})-
\d U_{\a}({\bar t})
+2i\sum_{\a'}\g_{\a'}\frac{\d U_{0}({\bar t})-
\d U_{\a'}({\bar t})}{\w-\W_{0}^{\ast}}\right]\right\}
\;\;
\label{sfwkw}
\end{eqnarray}
where $\W_{0}=\ve_{0}-i\g$. In the special case $\ve_{0}=0$, 
$\d U_{0}(t)=0$ and $\d U_{\a}(t)=\d U_{\a}=$ const, $\d J_{\a}(t)$ 
reduces to the time-dependent current in 
Eq. (\ref{DJFNFCO}) to first order in $\d U_{\a}$, as it should. 

Eq. (\ref{sfwkw}) takes a very simple form if $\ve_{0}=
\d U_{0}=\d U_{R}=0$ and $\a=R$: 
\begin{eqnarray}
\d J_{R}(t)&=&4e\g_{R}\g_{L}\int_{0}^{t}d{\bar t} \;
\d U_{L}(t-{\bar t}) 
\nonumber \\ &&
\times\left\{
\frac{\Im[f(i\g)]}{\g}{\rm e}^{-2\g {\bar t}}
-\frac{2}{\b}\Re\left[\sum_{n=0}^{\inf}
\frac{{\rm e}^{(i\w_{n}-\g){\bar t}}}{\w_{n}^{2}+\g^{2}}
\right]\right\},
\nonumber 
\end{eqnarray}
where $\w_{n}=(2n+1)\p i/\b+\m$ are the Matzubara frequencies and 
the identity 
$$
\int_{-\inf}^{\inf}\frac{d\w}{2\p}f(\w)
\frac{{\rm e}^{i\w {\bar t}}}{\w^{2}+\g^{2}}=
f(i\g)\frac{{\rm e}^{-\g {\bar t}}}{2\g}-
\frac{i}{\b}\sum_{n=0}^{\inf}
\frac{{\rm e}^{i\w_{n}{\bar t}}}{\w_{n}^{2}+\g^{2}},
$$
has been used. In the special case of a vanishing chemical potential, 
the Matzubara frequencies are imaginary numbers 
and $\d J_{R}(t)$ simplifies
\begin{eqnarray}
\d J_{R}(t)=
-2e\frac{\g_{R}\g_{L}}{\g}\int_{0}^{t}d{\bar t} \;
\d U_{L}(t-{\bar t}){\rm e}^{-2\g {\bar t}}
\nonumber \\ \times
\left\{\tan[\frac{\b\g}{2}]+
\frac{{\rm e}^{-[\frac{\p}{\b}-\g]{\bar t}}}{\p}
F({\bar t})
\right\},
\label{djmfkf}
\end{eqnarray}
where 
$$
F({\bar t})=\sum_{m=0,1}
(-)^{m}
\F[{\rm e}^{-\frac{2\p{\bar t}}{\b}},1,\frac{\p+(-)^{m}\b\g}{2\p}]
$$
is a linear combination of the Lerch transcendent functions
$\F[z,s,a]=\sum_{n=0}^{\inf}z^{n}/(a+n)^{s}$.

\begin{figure}[htbp]
\begin{center}
\includegraphics*[scale=0.8]{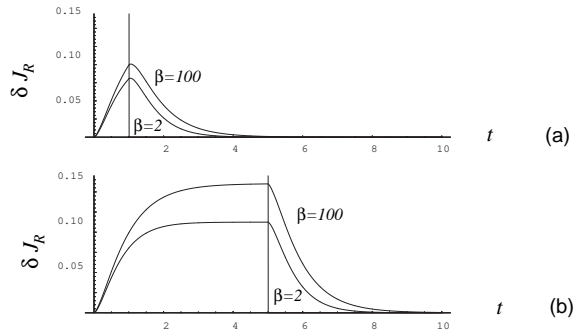}
\caption{{\footnotesize Current versus time for two different 
    external disturbance $\d U_{L}$. In both figures 
    $\g_{R}=\g_{L}=0.5$; the current is plotted for two different 
    inverse temperature $\b=2$ and $\b=100$. In (a) $\d U_{L}$ is a 
    square bump-like modulation whose duration is 1 while in (b) the 
    duration is 5.}}
\label{lcurrent_sb}
\end{center}
\end{figure}

In Fig. \ref{lcurrent_sb} we  
show  the trend of $\d J_{R}(t)$ for square bump-like 
modulations. On the top $\d U_{L}(t)=\Theta(t)\Theta(1-t)$ while on the 
bottom $\d U_{L}(t)=\Theta(t)\Theta(5-t)$; both disturbances are considered for two 
different inverse temperature $\b=2$ and $\b=100$. As one can see the 
effect of an increasing temperature consists in a sort of rescaling of 
the time-dependent current. The line widths $\g_{\a}$ have 
been taken equal and large enough to justify the linear approximation. 
Since the disturbance is of order 1, from Fig. \ref{stationary} one 
can see that $\g_{R}=\g_{L}=0.5$ is a good choice. 

\begin{figure}[htbp]
\begin{center}
\includegraphics*[scale=0.55]{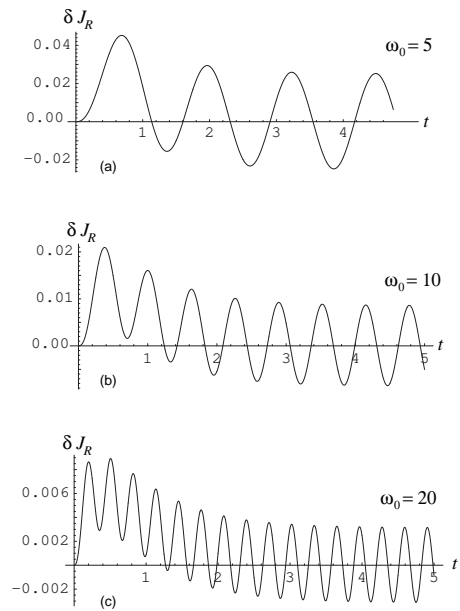}
\caption{{\footnotesize Current versus time for an oscillating 
    external disturbance $\d U_{L}$. In both figures 
    $\g_{R}=\g_{L}=0.5$ and the  
    inverse temperature is $\b=100$.  
    $\d U_{L}(t)=\sin\w_{0}t$ with $\w_{0}=5,\;10,\;20$ in  
    (a), (b) and (c) respectively.}}
\label{lcurrent_ac}
\end{center}
\end{figure}

\begin{figure}[htbp]
\begin{center}
\includegraphics*[scale=0.95]{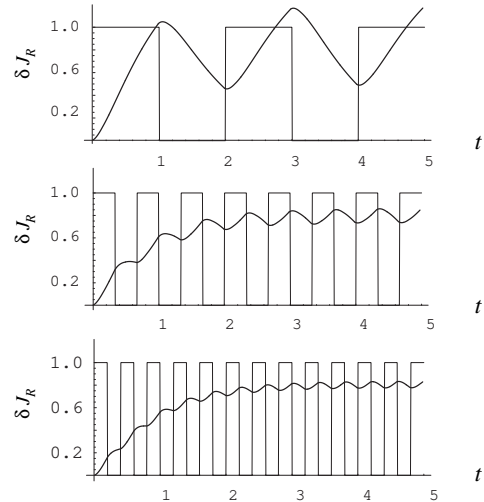}
\caption{{\footnotesize Current versus time for three different 
    periodic square bump-like modulations. In any figure
    $\g_{R}=\g_{L}=0.5$ and $\b=100$. The thin lines represent $\d 
    U_{L}$ while the thick lines represent $\d J_{R}$.}}
\label{steppe}
\end{center}
\end{figure}

The ac current in 
the linear approximation is plotted in Fig. \ref{lcurrent_ac} for 
$\b=100$ and $\g_{R}=\g_{L}=0.5$. The time-dependent disturbance is 
taken to be $\d U_{L}(t)=\sin\w_{0}t$ with $\w_{0}=5,\;10,\;20$ in 
(a), (b) and (c) respectively. Finally, in Fig. \ref{steppe} we have 
considered the current response to a periodic square bump-like 
modulation for different values of the period.

\section{Interacting Systems}
\label{interacting}

In earlier theoretical works on quantum transport one can distinguish
at least two schools. In one school one tries to keep the full
atomistic structure of the conductor and the leads, but
all works so far are at the level of the Local Density Approximation 
(LDA) and only the steady state has been considered. The advantage
of this approach is that the interaction in the leads and in the
conductor are treated on the same footing via self-consistent
calculations on the current-carrying system. It also allows for
detailed studies of how the contacts influence the conductance
properties. 

The other school is using simplified models which allows the
analysis to be carried much further. 
Considerable progresses have been made in this respect for a localized
level described by a Lundquist-like
model\cite{kral,lundin,xin} and for the so called ``Coulomb
island''\cite{meir,you} where $H_{0}$ in Eq. (\ref{qham}) is
replaced by the Anderson Hamiltonian. However, all these works
treat the leads as non-interacting, which prohibits a realistic
description of the contacts and of the long-range aspects of
the Coulomb interaction\cite{Büttiker}. The model approach are based on a
partitioned scheme which makes the time-dependent results
difficult to interpret. 

We here want to show how the current
LDA by Lang \textit{et al.}\cite{Lang1,Lang2} 
follows from the TDDFT scheme described in Section \ref{mftddft}.
We also present an exact result for the steady-state current 
of an interacting resonant tunneling system. Finally, the transient 
behavior of a capacitor-device-capacitor system is investigated 
on the level of Mean Field. 

\subsection{Steady-State Limit of TDDFT}

In Section \ref{model} we showed that under certain conditions
a steady state is reached in the long-time limit, and that
this limit is independent of history. We also showed that
the partitioned and partition-free treatments give an equivalent
description of the steady state. The mechanism for the 
loss of memory was pure dephasing, and it holds provided
the leads are macroscopic while the device is finite.
Another important ingredient is that the applied bias
is uniform deep inside the leads. With these assumptions,
our results can be generalized also to more general cases
than the simplified model explicitly considered in Section \ref{model}.
In TDDFT, the full interacting problem is reduced to a fictitious 
noninteracting one and \textit{ all} the results of Section \ref{model} 
can be recycled. In the case of Time Dependent Local Density 
Approximation (TDLDA), the exchange-correlation potential $v_{\rm xc}$ 
depends only on the instantaneous local density and has no memory at all. 
If the density tends to a constant, so does the effective potential 
$U^{\rm eff}$, which again 
implies that the density tends to a constant. Owing to the non-linearity 
of the problem there might still be more than one steady-state solution
or none at all.

If a steady state is reached in TDDFT, we can go directly to
the long-time limit of the Dyson equation and work in the frequency 
space. We may with no restriction
use a partitioned approach and split the fictitious one-electron Hamiltonian
matrix in a non-conducting part $\bcalE$ and a correction $\bV$ 
involving one-body hopping terms between the two leads and the device.
The lesser Green function of TDDFT fulfills 
$$
\bcalG^{<}(\ve)= [ 1 + \bcalG^{\rm R}(\ve) \bV ]\; \bg^<(\ve) 
[  1 + \bV \bcalG^{\rm A}(\ve) ] ,
$$
where $\bg$ is the uncontacted TDDFT Green function
[\textit{cf.} Eq. (\ref{newder3})]. In direct space, the uncontacted 
$\bg^<$ can be written
$$
g^<({\bf r}, {\bf r}', \ve) =
2 \pi i \sum_{m\alpha} f_\alpha(e_{m\alpha})
\phi_{m\alpha}({\bf r}) \phi_{m\alpha}^*({\bf r}')
\d(\ve-e_{m\alpha})
$$
in terms of diagonalizing orbitals $\phi_{m\alpha}$ 
with fictitious eigenvalues $e_{m\alpha}$ 
for the left and right leads ($\alpha = L,R$) and the device ($\alpha = D$)
and Fermi functions $f_\alpha$ with chemical potential $\mu_\alpha$.
The chemical potentials for the two leads differ, and the final result
is independent of the chosen chemical potential for the device.
When we apply $1 + \bcalG^{\rm R} \bV=
\bcalG^{\rm R}[\bg^{\rm R}]^{-1}$ to an unperturbed orbital 
$\phi_{m\alpha}$, it is transformed to an interacting, \textit{i.e.}, 
contacted
eigenstate $\psi_{m\alpha}$. Above the conductance threshold, 
states originating from the left lead become right-going scattering
states, and states from the right lead become left-going scattering
states. In addition, fully reflected waves and discrete state may arise
which contribute to the density but not to the current.
Thus,
$$
{\cal G}^<({\bf r}, {\bf r}', \ve) =
2 \pi i \sum_{m\alpha} f_\alpha(e_{m \alpha})
\psi_{m\alpha}({\bf r}) \psi_{m\alpha}^*({\bf r}')
\d(\ve-e_{m\alpha}).
$$
These results correspond closely to the general approach by
Lang and coworkers.\cite{Lang1,Lang2} 
In their approach, the continuum
is split into left and right-going parts, which are populated
according to two different chemical potentials. The density
is then calculated self-consistently. Lang \textit{et al.} further
approximate exchange and correlation by the LDA and the leads
by homogeneous jellia, but apart from these approximations it is
clear that his method implements TDDFT, as described
in Section \ref{mftddft}, in the steady state. 
It is also clear that the correctness of
Lang's approach relies on the Theorem of Equivalence between the 
partitioned and partition-free approaches and the Memory Loss Theorem
derived here. The equivalence between the scattering state
approach by Lang \textit{et al.} and the partitioned non-equilibrium
approach used by Taylor \textit{et al.}\cite{Taylor1,Taylor2}
 has also been shown by 
Brandbyge \textit{et al.}\cite{brand}

As shown above, the steady state of TDDFT can always be formulated in terms
of orbitals which diagonalize the asymptotic one-particle
Hamiltonian matrix. The current-carrying orbitals can always be 
grouped into a right-going class and a left-going class. 
As a consequence, the current can be expressed in a Landauer formula
\begin{equation}
J_{R}^{(\rm S)} =- e \sum_m [ f_L(e_{mL}) {\cal T}_{mL} - 
f_R(e_{mR}) {\cal T}_{mR} ]
\label{kefrv}
\end{equation}
in terms of fictitious transmission coefficients ${\cal T}_{m\a}$
and energy eigenvalues $e_{m\a}$, $\a=L,R$. 
We also wish to emphasize that the steady-state current in 
Eq. (\ref{kefrv}) comes out from a pure dephasing mechanism in the 
fictitious noninteracting problem. 
The memory-loss effects from scatterings is described by 
$A_{\rm xc}$ and $v_{\rm xc}$.

\subsection{One-Level Resonant Tunneling System}
 
In this Section we consider a 
resonant tunneling system described by the quadratic Hamiltonian 
of Eq. (\ref{qham}) and an inter-particle interaction 
$$
H_{W}=\frac{1}{2}\sum_{m\neq n}W_{m,n}n_{m}n_{n},
$$
where $n_{m}=c^{\dag}_{m}c_{m}$ is the occupation number operator of 
the level $m$ and $W_{m,n}=W_{n,m}$ is a symmetric matrix. (If $H_{W}$ 
includes long-range terms, the regrouping of potential terms as 
discussed in Section \ref {mftddft} must be done.) In the generalized 
TDDFT scheme (based on the $n_{m}$ occupations rather than on 
density) outlined in Section \ref{mftddft} the fictitious Green 
function ${\cal G}_{m,n}$ is obtained by solving the Dyson 
equations with $\bK=\bT+\bU^{\rm eff}$, where 
$$
U^{\rm eff}_{m,n}(t)=\d_{m,n}[U_{m}(t)+V_{{\rm H},m}(t)+v_{{\rm 
xc},m}(t)].
$$
If $\bU^{\rm eff}$ satisfies the 
hypothesis of the Theorem of Equivalence and of the Memory Loss 
Theorem we can use Eq. (\ref{stacg}) and 
write an exact formula for the steady-state current of 
an \textit{ interacting} resonant tunneling system:
\begin{eqnarray}
J^{(\rm S)}_{R}&=&-e\int\frac{d\ve}{2\p}f(\ve)
\label{stacgdft} \\ 
&&\times\left\{\G^{(0)}_{L}(\ve)\G_{R}(\ve+U^{\rm eff}_{L}(\ve))
\left|{\cal G}^{\rm R}_{0,0}(\ve+U^{\rm eff}_{L}(\ve))
\right|^{2}\right.
\nonumber \\ &&\left.\;\;+
\G^{(0)}_{R}(\ve)\G_{L}(\ve+U^{\rm eff}_{R}(\ve))
\left|{\cal G}^{\rm R}_{0,0}(\ve+U^{\rm eff}_{R}(\ve))
\right|^{2}
\right\}.
\nonumber
\end{eqnarray}
For normal-metal electrodes we expect that the effective potential 
$U^{\rm eff}_{\a}(\ve,t)\ra U^{\rm eff}_{\a}(\ve)={\rm const}$
provided $U_{\a}(\ve,t)\ra U_{\a}(\ve)={\rm const}$ when $t\ra\inf$. 
The constant $U^{\rm eff}_{\a}(\ve)$ may depend on the history of 
$U_{\a}(\ve,t)$ while the steady-state current is independent of the 
history of $U^{\rm eff}_{\a}(\ve,t)$. ${\cal G}^{\rm R}_{0,0}(\w)$ 
is given by Eq.(\ref{g00}) with $\tilde{\ve}_{0}=\ve_{0}+
\lim_{t\ra\inf}U^{\rm eff}_{0}(t)$ and with $\S^{\rm R}$ from 
Eq.(\ref{ser}) with $\tilde{\ve}_{k\a}=\ve_{k\a}+\lim_{t\ra\inf}
U^{\rm eff}_{k\a}(t)$. For the sake of clarity,  
Eq. (\ref{stacgdft}) has been written for systems having 
a one-to-one correspondence between the 
one-body indices $k\a$ and the one-body energies 
$\ve_{k\a}+U^{\rm eff}_{k\a}(0)$. The generalization to systems with 
degenerate levels is straightforward and it is left to the reader.

As a further example we study the RPA time-dependent current response in 
the partition-free approach. 
In the Hartree approximation the Green function 
$\bG^{\rm H}$ satisfies the equation of motion 
(\ref{ieom1td}) with $\bS_{c}=0$ and 
$$
\S^{\d}_{m,n}(z)\equiv \S^{\rm H}_{m,n}(z)=
\d_{m,n}\sum_{l:\;l\neq n}W_{n,l}\;n^{\rm H}_{l}(z),
$$
where $n^{\rm H}_{l}(z)=-iG^{\rm H,<}_{l,l}(z;z)$. According with the 
results obtained in Section \ref{keldyshth}, the lesser Green 
function $\bG^{\rm H,<}$ is given by Eq. (\ref{newder11}) with $\bG\ra 
\bG^{\rm H}$. Therefore, in the linear approximation we have
\begin{eqnarray}
&&\d \bG^{<}(t;t)\label{dief} \\ &&
\quad =-i\int d{\bar t}\; 
\bG_{0}^{\rm H,R}(t;{\bar t})[\d \bU^{\rm eff}({\bar t}),\bG^{\rm H,<}(0;0)]
\bG_{0}^{\rm H,A}({\bar t};t),
\nonumber
\end{eqnarray}
with
\begin{equation}
\d \bU^{\rm eff}(t)=\d \bU(t)+\d\bS^{\rm H}(t).
\label{efc;}
\end{equation}
Eqs. (\ref{dief})-(\ref{efc;}) form a coupled system of integral 
equations for the unknowns $\d \bG^{<}(t;t)$ and $\d \bU^{\rm eff}(t)$. 
For a capacitor-device-capacitor system one can take
$$
W_{m,n}=\left\{
\begin{array}{ll}
    W_{\a\a'} & {\rm if}\quad m=k\a,\; n=k'\a' \\
    0 & {\rm otherwise}
\end{array}
\right..
$$
Thus, putting an extra particle in the isolated $\a$ capacitor costs an 
energy $W_{\a\a}$ per particle. This means that the transfer of a 
finite number of particles from one capacitor to the other causes a finite 
change of the effective applied bias. We expect that the 
current vanishes in the long-time limit unless the applied bias 
continues to grow up. The coefficients $W_{RL}=W_{LR}$ mimic the 
repulsion energy between two particles in different capacitors.
Actually, one can also consider 
the interaction between a particle in the central device and another 
in one of the two capacitors. No extra complications arise if 
$W_{0,k\a}=W_{\a}$, $\forall k$, and the results we are going to obtain 
can be easily extended. 

Switching a bias $\d U_{k\a}(t)=\d U_{\a}(t)$,  
from Eq. (\ref{efc;}) one gets 
$\d U^{\rm eff}_{m,n}(t)=\d_{m,n}\d U^{\rm eff}_{n}(t)$ 
with $\d U^{\rm eff}_{k\a}(t)=\d U^{\rm eff}_{\a}(t)$, $\forall k$, 
and 
\begin{equation}
\d U^{\rm eff}_{\a}(t)=\d U_{\a}(t)-
\frac{1}{e}\sum_{\b}\int_{0}^{t}d\bar{t}\;W_{\a\b}\d J_{\b}(\bar{t}),
\label{dur}
\end{equation}
where it has been taken into account that $\d N^{\rm H}_{\a}(t)\equiv \sum_{k}
\d n^{\rm H}_{k\a}(t)=-\frac{1}{e}
\int_{0}^{t}d\bar{t}\;\d J_{\a}(\bar{t})$.
Since $\d \bU^{\rm eff}$ has the same matrix structure of the bare 
$\d \bU$, in the wide-band limit the linear time-dependent current 
$\d J_{\a}(t)$ is given by Eq. (\ref{sfwkw}) with $\d \bU$ replaced by 
$\d \bU^{\rm eff}$. (It is worth noticing that 
the wide band limit still makes sense if the line width  
is approximately constant in a small interval around 
the chemical potential $\m$.) 
In this way the 
system of Eqs. (\ref{dief})-(\ref{efc;}) is reduced to a system of 4  
coupled integral equations for the 4 scalar unknowns  
$\d U^{\rm eff}_{\a}$, $\d J_{\a}$ with $\a=L,R$. The symmetric 
case $\g_{R}=\g_{L}=\g/2$, $W_{RR}=W_{LL}$ allows a further 
simplification. Let us define $\d U^{\rm eff}_{\pm}=
\d U^{\rm eff}_{R}\pm \d U^{\rm eff}_{L}$, $\d U_{\pm}=
\d U_{R}\pm \d U_{L}$, 
$\d J_{\pm}=\d J_{R}\pm \d J_{L}$ and $W_{\pm}=W_{RR}\pm W_{RL}=
W_{LL}\pm W_{LR}$. Then, 
from Eq. (\ref{dur}) we find
\begin{equation}
\d U^{\rm eff}_{\pm}(t)=\d U_{\pm}(t)-
\frac{W_{\pm}}{e}\int_{0}^{t}d\bar{t}\;\d J_{\pm}(\bar{t}),
\label{dutot}
\end{equation}
while from Eq. (\ref{sfwkw})
\begin{equation}
\d J_{+}(t)=2e\g\int_{0}^{t}d{\bar t} \;
C_{+}(t-{\bar t})[2\d U_{0}({\bar t})-\d U^{\rm eff}_{+}({\bar t})],
\label{sieklefl}
\end{equation}
\begin{equation}
\d J_{-}(t)=-2e\g\int_{0}^{t}d{\bar t} \;
C_{-}(t-{\bar t})\d U^{\rm eff}_{-}({\bar t}),
\label{didke}
\end{equation}
where
$$
C_{\pm}(t)=\Re\left\{\int\frac{d\w}{2\p}f(\w)
\frac{{\rm e}^{i(\w-\ve_{0}+i\g)t}}{\w-\ve_{0}\mp i\g}
\right\}
$$
is the conductivity kernel. Once $\d J_{\pm}$ has been obtained, one can 
calculate $\d J_{R}=(\d J_{+}+\d J_{-})/2$ and 
$\d J_{L}=(\d J_{+}-\d J_{-})/2$. 
\begin{figure}[htbp]
\begin{center}
\includegraphics*[scale=0.6]{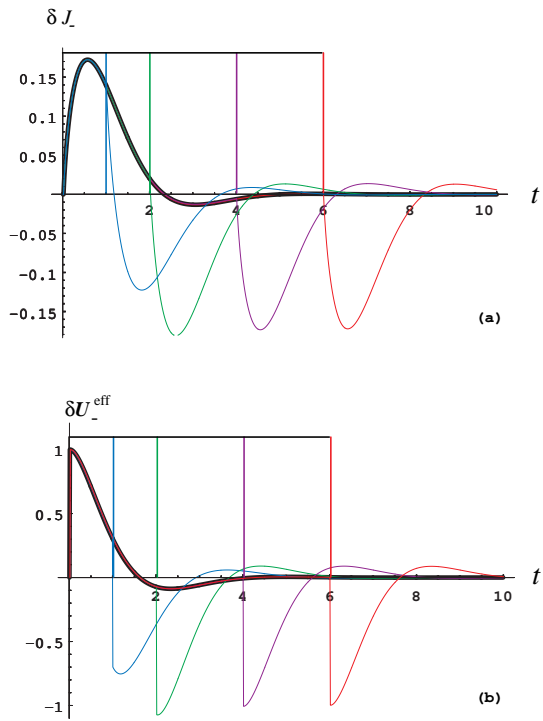}
\caption{{\footnotesize Numerical solutions of  
Eqs. (\ref{dutot})-(\ref{sieklefl})-(\ref{didke}) in the zero temperature limit with 
$\m=\ve_{0}=0$, $W_{-}=5$ and an external disturbance as described in the main 
text. The thick lines are the current in (a) and the effective potential 
in (b) for the step like modulation. 
The currents and the effective potentials 
for $t_{0}=1,\;2,\;4,\;6$ unstick from the thick line and start to 
oscillate and eventually vanish after a time $t\propto 1/W_{-}$. 
The vertical lines 
are the bare applied potentials.}}
\label{plasmonjdc}
\end{center}
\end{figure}

In order to illustrate what is the time-dependent response of this 
model we have considered the zero temperature case with $\d 
U_{R}(t)=-\d U_{L}(t)=(1/2)\Theta(t)\Theta(t_{0}-t)$ and $\d U_{0}(t)=0$. 
Then, $\d U_{+}(t)=0$ and hence $\d J_{+}(t)=\d U^{\rm 
eff}_{+}(t)=0$. It follows that $\d J_{R}(t)=-\d J_{L}(t)=\d 
J_{-}(t)/2$ and $\d U^{\rm eff}_{R}(t)=-\d U^{\rm eff}_{L}(t)=
\d U^{\rm eff}_{-}(t)/2$ for any time $t$. In 
Fig. \ref{plasmonjdc}(a) we display the time-dependent current for 
square bump-like modulations with $t_{0}=1,\;2,\;4,\;6$ and 
$W_{-}=5$. The thick line is the current for the step like 
modulation $\d U_{-}(t)=\Theta(t)$; 
depending on the value of $t_{0}$ the current unsticks itself from 
the thick line giving rise to different damped oscillating curves. In 
correspondence of each $t_{0}$ a vertical line has been drawn; it 
represents the bare applied potential 
$\d U_{-}(t)$. Fig. \ref{plasmonjdc}(b)  shows the 
time-dependent effective potential $\d U^{\rm eff}_{-}(t)$. 
As the current response, it drops to zero in the long-time limit since the 
interactions completely screen the applied bias after a time 
$t\propto 1/W_{-}$.

\section{Summary and Concluding Remarks}
\label{conclusion}

In the present work we have used a partition-free scheme  
in order to treat the time-dependent current response of a 
mesoscopic system coupled to macroscopic leads. 
To this end, we have further developed the Keldysh formalism and 
we have formulated a formally exact theory which is more akin to the 
way the experiments are carried out. Among the advantages of the 
partition-free scheme we stress the possibility to calculate physical
dynamical responses and to include the interactions between 
the leads and between the leads and the device in a quite natural way. 

In the noninteracting case we have shown that a perfect destructive 
interference takes place provided the energy levels of the leads form 
a continuum. The steady-state develops due to a \textit{dephasing mechanism}.
The comparison of our steady-state current with that obtained in 
the partitioned scheme shows that the two currents are equivalent if the energy 
levels are properly shifted in order to preserve charge neutrality. 
This kind of equivalence remains true for any time-dependent external 
potentials (Theorem of Equivalence). 
The Theorem of Equivalence has then been used in order to 
prove that the steady-state current depends 
only on the asymptotic value of the external 
perturbation (Memory Loss Theorem). For the sake of clarity, 
the Theorem of Equivalence and the Memory Loss Theorem have been 
proved for a single-level central device. The generalization to a 
multi-level central device is straightforward, as can be readily 
verified.
In the wide band limit we have obtained an analytic result 
for the time-dependent current in the case of a step-like modulation 
and for arbitrary modulations in the linear regime. 

The interacting case represents a more difficult challenge and 
the expression for the lesser Green function at any finite time is 
more complicated than that commonly used to calculate steady-state 
response functions. 
As an alternative to a full many-body treatment we have
proposed a formally exact one-particle scheme based on TDDFT. 
Then, \textit{all} the results obtained in the noninteracting case can be 
recycled provided we substitute the external potential with the exact 
effective potential of TDDFT. 
Although it is difficult to prove any rigorous results
for the effective TDDFT potential, we expect 
the interactions to reduce the memory effects even further
compared to the noninteracting case. Thus, 
any nonlinear steady-state current can been expressed in a 
Landauer-like formula in terms of 
fictitious transmission coefficients and one-particle energy 
eigenvalues. The steady-state current depends on history only through 
the asymptotic shape of the effective TDDFT potential.
This exact result 
may prompt for new approximations to the exchange-correlation 
action functional $A_{\rm xc}$. 
In the effective one-particle scheme of TDDFT the steady-state 
current comes out from a pure dephasing mechanism. The damping 
mechanism (due to the electron-electron scatterings) of the real 
problem is described by $A_{\rm xc}$.
As an illustrative example we  
have also calculated 
the RPA time-dependent current of a capacitor-device-capacitor 
system and we have displayed the effect of the charge
oscillations in the discharge process.

\begin{acknowledgments}
We would like to acknowledge useful discussions with
U. von Barth, 
P. Bokes, 
M. Cini, R. Godby, A.-P. Jahuo, B. I. Lundqvist, P. Hyldgaard,
and B. Tobiyaszewska.
This work was supported by the RTN program of the European Union
NANOPHASE (contract HPRN-CT-2000-00167).
\end{acknowledgments}

\appendix

\section{Proof of Eq. (\ref{GGGG})}
\label{z}

It is convenient to define $\bG_{0}$ as the 
solution of Eqs. (\ref{ieom1td}) with $\bS_{\rm c}=0$. 
$\bG_{0}$ satisfies all the relations we have derived for a 
noninteracting system in the presence of an external disturbance. By 
using the Langreth theorem, we get
\begin{eqnarray}
\bG^{\lessgtr}&=&\left[\d+
\bG^{\rm R}\cdot\bS^{\rm R}\right]\cdot \bG_{0}^{\lessgtr}+
\bG^{\lessgtr}\cdot\bS^{\rm A}\cdot \bG_{0}^{\rm A}
\nonumber \\ &&+
\left[\bG^{\rm R}\cdot\bS^{\lessgtr}+
\bG^{\rceil}\star\bS^{\lceil}\right]
\cdot \bG_{0}^{\rm A}
\nonumber \\ &&+
\bG^{\rm R}\cdot \bS^{\rceil}\star \bG_{0}^{\lceil}+
\bG^{\rceil}\star\bS\star \bG_{0}^{\lceil}
\nonumber
\end{eqnarray}
and solving for $\bG^{\lessgtr}$ 
\begin{eqnarray}
\bG^{\lessgtr}&=&[\d+\bG^{\rm R}\cdot\bS^{\rm R}]\cdot \bG_{0}^{\lessgtr}
\cdot [\d+\bS^{\rm A}\cdot \bG_{0}^{\rm A}]
\nonumber \\ 
&&+
\left[\bG^{\rm R}\cdot\bS^{\lessgtr}+
\bG^{\rceil}\star\bS^{\lceil}\right]\cdot \bG^{\rm A}
\nonumber \\ &&+
[\bG^{\rm R}\cdot \bS^{\rceil}\star \bG_{0}^{\lceil}
+\bG^{\rceil}\star\bS\star \bG_{0}^{\lceil}]\cdot
[\d+\bS^{\rm A}\cdot \bG_{0}^{\rm A}].
\nonumber
\end{eqnarray}
Next, we use
$$
\bG_{0}^{\lessgtr}(t;t')=\bG_{0}^{\rm R}(t;0)\bG_{0}^{\lessgtr}(0;0)
\bG_{0}^{\rm A}(0;t')
$$
and 
$$
\bG_{0}(\t;t)=-i\bG_{0}(\t;0)\bG_{0}^{\rm A}(0;t),
$$
so that
\begin{eqnarray}
&&\bG^{\lessgtr}(t;t')=\bG^{\rm R}(t;0)\bG_{0}^{\lessgtr}(0;0)
\bG^{\rm A}(0;t')
\label{tg<>}\\ &&\quad\quad
+\left[\bG^{\rm R}\cdot\bS^{\lessgtr}\cdot \bG^{\rm A}\right](t;t')+
\left[\bG^{\rceil}\star\bS^{\lceil}\cdot \bG^{\rm A}\right](t;t')
\nonumber \\ 
&&\quad\quad -i\left[
\bG^{\rm R}\cdot\bS^{\rceil}\star \bG_{0}+
\bG^{\rceil}\star \bS\star \bG_{0}
\right](t;0)\bG^{\rm A}(0;t').
\nonumber
\end{eqnarray}
As in the noninteracting case, we proceed by writing down the Dyson 
equation for $\bG(t;\t)$. Taking into account that 
\begin{equation}
\bG(\t;\t')=\bG_{0}(\t;\t')+\left[
\bG_{0}\star\bS\star \bG
\right](\t;\t')
\label{termg}
\end{equation}
and that 
$$
\bG_{0}(t;\t)=i\bG^{\rm R}_{0}(t;0)\bG_{0}(0;\t),
$$ 
we have
\begin{equation}
\bG(t;\t)=\left[
\bG^{\rm R}\cdot \bS^{\rceil}\star \bG
\right](t;\t)+i\bG^{\rm R}(t;0)\bG(0;\t).
\label{g-|}
\end{equation}
Similarly, it is straightforward to show that
\begin{equation}
\bG(\t;t)=\left[\bG\star \bS^{\lceil}\cdot 
\bG^{\rm A}
\right](\t;t)-i\bG(\t;0)\bG^{\rm A}(0;t).
\label{g|-}
\end{equation}
Substituting Eq. (\ref{g-|}) into Eq. (\ref{tg<>}) and using Eq. (\ref{termg}) 
one finds
\begin{eqnarray}
\bG^{\lessgtr}(t;t')&=&\bG^{\rm R}(t;0)\bG^{\lessgtr}(0;0)
\bG^{\rm A}(0;t')
\nonumber \\ && +
\left[\bG^{\rm R}\cdot\left[\bS^{\lessgtr}+
\bS^{\rceil}\star \bG\star \bS^{\lceil}
\right]\cdot \bG^{\rm A}\right](t;t')
\nonumber \\
&&+i\bG^{\rm R}(t;0)\left[
\bG\star \bS^{\lceil}\cdot \bG^{\rm A}
\right](0;t')\nonumber \\ &&
-i\left[
\bG^{\rm R}\cdot\bS^{\rceil}\star \bG
\right](t;0)\bG^{\rm A}(0;t').
\label{tg<>3}
\end{eqnarray}
Using Eqs. (\ref{g-|})-(\ref{g|-}) to express the last two 
terms as
\begin{eqnarray}
i\bG^{\rm R}(t;0)\left[
\bG\star \bS^{\lceil}\cdot \bG^{\rm A}
\right](0;t')
\quad\quad\quad\quad\quad
\quad\quad\quad\quad
\nonumber \\ =
i\bG^{\rm R}(t;0)\bG^{>}(0;t')-
\bG^{\rm R}(t;0)\bG^{>}(0;0)\bG^{\rm A}(0;t')
\nonumber 
\end{eqnarray}
and 
\begin{eqnarray}
-i\left[
\bG^{\rm R}\cdot\bS^{\rceil}\star \bG
\right](t;0)\bG^{\rm A}(0;t')
\quad\quad\quad\quad
\quad\quad\quad\quad\quad
\nonumber \\ 
=-i\bG^{<}(t;0)\bG^{\rm A}(0;t')-
\bG^{\rm R}(t;0)\bG^{<}(0;0)\bG^{\rm A}(0;t'),
\nonumber
\end{eqnarray}
we end up with Eq. (\ref{GGGG}).  

\section{Proof of Eq. (\ref{INTERM})}
\label{a1}

Due to the smoothness of the self energy, in the long-time limit we 
can use the Riemann-Lebesgue theorem to obtain the following 
asymptotic behaviors 
\begin{equation}
\lim_{t\ra\inf}G^{\rm R}_{0,0}(t;0)=
\lim_{t\ra\inf}\sum_{k}G^{\rm A}_{0,k\a}(0;t)V_{k\a}=0
\label{asymg00}
\end{equation}
and
$$
\lim_{t\ra\inf}G^{\rm R}_{0,k\a}(t;0)=
-iV_{k\a}{\rm e}^{-i\tilde{\ve}_{k\a}t}G^{\rm R}_{0,0}(\tilde{\ve}_{k\a}).
$$
\begin{eqnarray}
\lim_{t\ra\inf}\sum_{k}G^{\rm A}_{k'\a',k\a}(0;t)V_{k\a}=
iV_{k'\a'}{\rm e}^{i\tilde{\ve}_{k'\a'}t}
\quad\quad
\nonumber \\
\times\left[
\d_{\a,\a'}+G^{\rm A}_{0,0}(\tilde{\ve}_{k'\a'})
\S^{\rm A}_{\a}(\tilde{\ve}_{k'\a'})
\right].
\nonumber
\end{eqnarray}
From the above results  
and the definition (\ref{qalfa}) one has 

\begin{eqnarray}
\lim_{t\ra\inf}Q_{\a}(\z;t)=
\sum_{k'}\frac{V^{2}_{k'\a}}{\z-\ve_{k'\a}}
G^{\rm R}_{0,0}(\tilde{\ve}_{k'\a})
\quad\quad\quad\quad
\label{statcurint} \\ \quad\quad
+\sum_{k'\a'}\frac{V^{2}_{k'\a'}}{\z-\ve_{k'\a'}}
G^{\rm R}_{0,0}(\tilde{\ve}_{k'\a'})
G^{\rm A}_{0,0}(\tilde{\ve}_{k'\a'})
\S^{\rm A}_{\a}(\tilde{\ve}_{k'\a'})
\nonumber \\ \quad\quad+
\lim_{t\ra\inf}G_{0,0}(\z)\sum_{k'\a'}\frac{V^{2}_{k'\a'}}
{\z-\ve_{k'\a'}}
G^{\rm R}_{0,0}(\tilde{\ve}_{k'\a'}){\rm e}^{-i\tilde{\ve}_{k'\a'}t}
\nonumber \\ \quad\quad
\quad\quad\quad\quad
\quad\quad\quad\quad\quad\;\;
\times\sum_{k''}\frac{V^{2}_{k''\a}}{\z-\ve_{k''\a}}
{\rm e}^{i\tilde{\ve}_{k''\a}t}
\nonumber \\ \quad\quad
+\lim_{t\ra\inf}G_{0,0}(\z)\sum_{k'\a'}\frac{V^{2}_{k'\a'}}
{\z-\ve_{k'\a'}}G^{\rm R}_{0,0}(\tilde{\ve}_{k'\a'})
{\rm e}^{-i\tilde{\ve}_{k'\a'}t} 
\nonumber \\ \quad\quad\times
\sum_{k''\a''}\frac{V^{2}_{k''\a''}}
{\z-\ve_{k''\a''}}
G^{\rm A}_{0,0}(\tilde{\ve}_{k''\a''})
\S^{\rm A}_{\a}(\tilde{\ve}_{k''\a''})
{\rm e}^{i\tilde{\ve}_{k''\a''}t}
\nonumber \\ \nonumber \\ =
\int\frac{d\ve}{2\p}\frac{\G_{\a}(\ve)}{\z-\ve+ U_{\a}}
G^{\rm R}_{0,0}(\ve)
\quad\quad\quad
\quad\quad\quad
\quad\quad\quad
\nonumber \\ \quad\quad+\sum_{\a'}\int\frac{d\ve}{2\p}
\frac{\G_{\a'}(\ve)}{\z-\ve+ U_{\a'}}
\left|G^{\rm R}_{0,0}(\ve)\right|^{2}
\S^{\rm A}_{\a}(\ve)
\nonumber \\ \quad\quad +
\lim_{t\ra\inf}G_{0,0}(\z)\int\frac{d\ve}{2\p}
\G_{\a}(\ve)\frac{{\rm e}^{i\ve t}}{\z-\ve+U_{\a}}
 \nonumber \\ \quad\quad\quad\quad\quad
\times\sum_{\a'}\int\frac{d\ve'}{2\p}\G_{\a'}(\ve')
\frac{{\rm e}^{-i\ve' t}}{\z-\ve'+U_{\a'}}
G^{\rm R}_{0,0}(\ve')
\nonumber \\ +
\lim_{t\ra\inf}G_{0,0}(\z)
\sum_{\a'}\int\frac{d\ve'}{2\p}\G_{\a'}(\ve')
G^{\rm R}_{0,0}(\ve')\frac{{\rm e}^{-i\ve' t}}{\z-\ve'+U_{\a'}}
 \nonumber \\ \times
\sum_{\a''}\int\frac{d\ve''}{2\p}\G_{\a''}(\ve'')
G^{\rm A}_{0,0}(\ve'')\S^{\rm A}_{\a}(\ve'')
\frac{{\rm e}^{i\ve'' t}}{\z-\ve''+U_{\a''}},
\nonumber
\end{eqnarray}
where the relation 
$$
G_{k\a,k'\a'}(\z)=\frac{\d_{k\a,k'\a'}}{\z-\ve_{k\a}}
+\frac{V_{k\a}}{\z-\ve_{k\a}}\frac{V_{k'\a'}}{\z-\ve_{k'\a'}}
G_{0,0}(\z),
$$
has been explicitly used. Since 
$\z\in\G$ the quantity $[\z-\ve-U_{\a}]^{-1}$ is a smooth function 
of $\ve$ for any real $\ve$. 
In the limit $t\ra\inf$ the last two terms in Eq. (\ref{statcurint}) 
vanish according with the Riemann-Lebesgue theorem and Eq. (\ref{INTERM}) 
is recovered.

\section{Proof of Eq. (\ref{DJFNFCO})}
\label{a2}

The quantity $Q_{\a}(\z;t)$ involves the multiplication of three matrices and we 
can recognize four contributions, two containing $G^{\rm R}_{0,0}$ 
and other two containing $G^{\rm R}_{0,k'\a'}$. It is 
straightforward to verify that 
\begin{equation}
G_{0,0}(\z)=
\left\{
\begin{array}{ll}
\frac{1}{\z+i\g} & \Im[\z]>0 \\ & \\
\frac{1}{\z-i\g} & \Im[\z]<0
\end{array}\right. ,
\label{tgooz}
\end{equation}
and that $G^{\rm R,A}_{0,0}(\w)=[\w\pm i\g]^{-1}$, where $\g=\g_{R}+\g_{L}$. Hence
\begin{eqnarray}
&& G^{\rm R}_{0,0}(t;0)=-i{\rm e}^{-\g t},
\label{az,`/} \\ 
&&G^{\rm R}_{0,k\a}(t;0)=
-iV_{k\a}\frac{{\rm e}^{-i\tilde{\ve}_{k\a}t}-{\rm e}^{-\g t}}
{\tilde{\ve}_{k\a}+i\g}
\nonumber \\ 
&& \sum_{k}V_{k\a}G^{\rm A}_{0,k\a}(0;t)=
-\g_{\a}{\rm e}^{-\g t},
\nonumber \\ 
&& 
\sum_{k'}V_{k'\a'}G^{\rm A}_{k\a,k'\a'}(0;t)=
i\d_{\a,\a'}V_{k\a}{\rm e}^{i\tilde{\ve}_{k\a}t}
\nonumber \\ &&
\quad\quad\quad\quad\quad
\quad\quad\quad\quad\quad\;\;
-\g_{\a'}V_{k\a}\frac{{\rm e}^{i\tilde{\ve}_{k\a}t}-{\rm e}^{-\g t}}
{\tilde{\ve}_{k\a}-i\g}.
\nonumber
\end{eqnarray}

\begin{figure}[htrbb]
\begin{center}
\includegraphics*[scale=0.75]{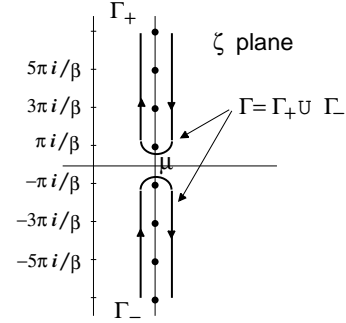}
\caption{{\footnotesize Contour $\G$ of Eq. (\ref{current2}). The black 
dots correspond to the position of the Matzubara frequencies in the 
complex $\z$ plane.}}
\label{lwcontour}
\end{center}
\end{figure}

Eqs. (\ref{tgooz})-(\ref{az,`/}) are all what we need in 
order to evaluate the quantity $Q_{\a}(\z;t)$ in Eq. (\ref{qalfa}). 
The time-dependent current is then obtained integrating 
$Q_{\a}(\z;t)f(\z){\rm e}^{\eta\z}$ over $\z$ along the contour $\G$ of 
Fig. \ref{lwcontour}, according with Eq. (\ref{current2}). 
Using Eqs. (\ref{az,`/}) and expressing $G_{k\a,0}(\z)=G_{0,k\a}(\z)$ 
and $G_{k\a,k'\a'}(\z)$ in terms of $G_{0,0}(\z)$ we obtain 
\begin{equation}
G^{\rm R}_{0,0}(t;0)G_{0,0}(\z)\sum_{k}V_{k\a}G^{\rm A}_{0,k\a}(0;t)
=i\g_{\a}G_{0,0}(\z){\rm e}^{-2\g t},
\label{tdcr1}
\end{equation}
\begin{eqnarray}
\sum_{k'\a'}G^{\rm R}_{0,0}(t;0)G_{0,k'\a'}(\z)
\sum_{k}V_{k\a}G^{\rm A}_{k'\a',k\a}(0;t)=
\quad\quad\quad
\label{tdcr2} \\ =
2\g_{\a}{\rm e}^{-\g t}G_{0,0}(\z)
\int\frac{d\ve}{2\p}\frac{{\rm e}^{i\ve t}}{\z-\ve+U_{\a}} +
\nonumber \\ 
2i\g_{\a}{\rm e}^{-\g t}G_{0,0}(\z)\sum_{\a'}\g_{\a'}
\int\frac{d\ve}{2\p}
\frac{{\rm e}^{i\ve t}-{\rm e}^{-\g t}}{(\z-\ve+U_{\a'})(\ve-i\g)}.
\nonumber 
\end{eqnarray}
We are left with the contributions containing $G^{\rm R}_{0,k'\a'}$. 
One of them is quite easy to evaluate and yields:

\begin{eqnarray}
\sum_{k'\a'}G^{\rm R}_{0,k'\a'}(t;0)G_{k'\a',0}(\z)
\sum_{k}V_{k\a}G^{\rm A}_{0,k\a}(0;t)=
\quad\quad\quad
\label{tdcr3} \\ =
2i\g_{\a}G_{0,0}(\z)\sum_{\a'}\g_{\a'}
\int\frac{d\ve}{2\p}
\frac{{\rm e}^{-i\ve t-\g t}-{\rm e}^{-2\g t}}{(\z-\ve+U_{\a'})(\ve+i\g)}.
\nonumber
\end{eqnarray}
The other one is much more involved, but nothing more than standard 
algebra is needed to get the following expression
\begin{eqnarray}
\sum_{k'\a'}\sum_{k''\a''}
G^{\rm R}_{0,k'\a'}(t;0)G_{k'\a',k''\a''}(\z)
\times
\quad\quad\quad\quad\quad
\quad\quad
\label{tdcr5} \\ 
\sum_{k}V_{k\a}G^{\rm A}_{k''\a'',k\a}(0;t)=
\nonumber \\
=2\g_{\a}\int\frac{d\ve}{2\p}
\frac{1-{\rm e}^{i\ve t}{\rm e}^{-\g t}}
{(\ve+i\g)(\z-\ve+U_{\a})}+
\quad\quad\quad\quad\quad
\quad\quad
\nonumber \\ 
2i\g_{\a}\sum_{\a'}\g_{\a'}\int\frac{d\ve}{2\p}
\frac{1}{\z-\ve+U_{\a'}}\left|
\frac{{\rm e}^{i\ve t}-{\rm e}^{-\g t}}
{\ve+i\g}\right|^{2}-
\nonumber \\ 
2iG_{0,0}(\z)\sum_{\a'}\g_{\a'}\int\frac{d\ve}{2\p}
\frac{{\rm e}^{-i\ve t}-{\rm e}^{-\g t}}
{(\ve+i\g)(\z-\ve+U_{\a'})}\int\frac{d\ve'}{2\p}
\times
\nonumber \\ 
\left[
2i\g_{\a}
\frac{{\rm e}^{i\ve' t}}{\z-\ve' +U_{\a}}-
\frac{{\rm e}^{i\ve' t}-{\rm e}^{-\g t}}
{(\ve'-i\g)(\z-\ve'+U_{\a''})}
\right].
\nonumber
\end{eqnarray}
The r.h.s. of the above four equations must now be multiplied by 
$f(\z){\rm e}^{\eta\z}$ and integrated over $\z$ along the contour $\G$. 
Smearing the branches $\G_{+}$ and $\G_{-}$ 
on the real axis and taking into account Eq. (\ref{tgooz}), the 
r.h.s. of Eq. (\ref{tdcr1}) yields the following contribution to the 
current
\begin{equation}
4e\g_{\a}{\rm e}^{-2\g t}\;\Im\left\{
\int\frac{d\w}{2\p}f(\w)\frac{1}{\w+i\g}
\right\},
\label{ctdcn1}
\end{equation}
where the integration over $\w$ has to be understood from 
$-\inf$ to $+\inf$. 
Another contribution comes from the first term on the r.h.s. of Eq. (\ref{tdcr2}). 
By closing the contour of the $\ve$ integration on the complex upper 
half plane, it is non vanishing only if ${\rm 
Im}[\z]>0$. Therefore, only the upper branch $\G_{+}$ of $\G$ 
contributes. $\G_{+}$ can then be smeared on the real axis and one 
gets
\begin{equation}
-4e\g_{\a}{\rm e}^{-\g t}\;\Im\left\{
\int\frac{d\w}{2\p}f(\w)\frac{{\rm e}^{i(\w+U_{\a})t}}{\w+i\g}
\right\}.
\label{ctdcn2}   
\end{equation}
A similar procedure can be adopted to evaluate the contribution 
coming from the second term on the r.h.s. of Eq. (\ref{tdcr2}). 
One more time we can close the contour  of the $\ve$ integration on 
the complex upper half plane. The pole in $\ve=i\g$ does not contribute 
since its residue is zero. The other pole is in $\ve=\z+U_{\a'}$ and 
hence one obtains
\begin{equation}
-4e\g_{\a}\;\Re\left\{
\int\frac{d\w}{2\p}\frac{f(\w)}{\w+i\g}\sum_{\a'}\g_{\a'}
\frac{{\rm e}^{i(\w+U_{\a'}) t-\g t}-{\rm e}^{-2\g t}}{\w+U_{\a'}-i\g}\right\}.
\label{ctdcn3}
\end{equation}
Next, we have to calculate the contribution coming from Eq. (\ref{tdcr3}). 
By the same reasoning leading to Eq. (\ref{ctdcn3}) it is readily 
verified that it yields the same result. 
Therefore we have to keep in mind that Eq. (\ref{ctdcn3}) should be 
multiplied by 2 at the end. Let us now consider the contribution 
coming from the first two terms on the r.h.s. of Eq. (\ref{tdcr5}). 
Since the discontinuous function $G_{0,0}(\z)$ does not appear in the 
integrand we can perform the contour integral over $\z$. We find
\begin{eqnarray}
&-&4e\g_{\a}\;\Im\left\{\int\frac{d\w}{2\p}
f(\w-U_{\a})\frac{1-{\rm e}^{i\w t}{\rm e}^{-\g t}}
{\w+i\g}\right\}-
\label{ctdcn5} \\ 
&&4e\g_{\a}\;\Re\left\{
\sum_{\a'}\g_{\a'}\int\frac{d\w}{2\p}
f(\w-U_{\a'})
\left|
\frac{{\rm e}^{i\w t}-{\rm e}^{-\g t}}
{\w+i\g}\right|^{2}
\right\}.
\nonumber
\end{eqnarray}
The contribution coming from the last two terms on 
the r.h.s. of Eq. (\ref{tdcr5}) vanishes. Indeed the integral over $\ve$ can be 
closed on the complex lower half plane. The pole in $\ve=-i\g$ does not 
contribute since its residue is zero. The other pole contributes only 
if $\Im[\z]<0$. At the same time we can also perform the integration 
over $\ve'$ by closing the contour in the complex upper half plane. 
The first term in the square brackets of Eq. (\ref{tdcr5}) is non 
vanishing only if $\Im[\z]>0$. The same holds for the second 
term since the pole $\ve'=i\g$ has vanishing residue. By 
collecting all the results obtained one sees that they can be grouped 
into three broad categories: those  
which are time independent and that give rise to the stationary 
current, those which are proportional to ${\rm e}^{-\g t}$ and those 
which are proportional to ${\rm e}^{-2\g t}$. 
These last ones can be rewritten as
\begin{equation}
-4e\g_{\a}{\rm e}^{-2\g t}\int\frac{d\w}{2\p}f(\w)
\sum_{\a'}\frac{\g_{\a'}U^{2}_{\a'}}{[\w^{2}+\g^{2}][(\w+U_{\a'})^{2}+\g^{2}]}.
\label{e2gt}
\end{equation}
Let us now group the terms proportional to ${\rm e}^{-\g t}$. Two of them 
comes from  Eq. (\ref{ctdcn2}) and the first term of Eq. (\ref{ctdcn5}); 
their sum can be written as
\begin{equation}
-4e\g_{\a}U_{\a}{\rm e}^{-\g t}\int\frac{d\w}{2\p}f(\w)
\;\Im\left\{\frac{{\rm e}^{i(\w+U_{\a})t}}
{(\w+i\g)(\w+U_{\a}+i\g)}\right\}.
\label{egt1}
\end{equation}
The other two pieces come from Eq. (\ref{ctdcn3}) (which we recall must 
be multiplied by 2) and the last term of Eq. (\ref{ctdcn5}). By 
writing explicitly the real part, after some algebra one finds
\begin{eqnarray}
-8e\g_{\a}{\rm e}^{-\g t}\int\frac{d\w}{2\p}f(\w)
\sum_{\a'}\g_{\a'}U_{\a'}\times
\quad\quad\quad\quad
\quad\quad
\label{egt2} \\
\frac{\w\cos[(\w+U_{\a'})t]+\g\sin[(\w+U_{\a'})t]}
{[\w^{2}+\g^{2}][(\w+U_{\a'})^{2}+\g^{2}]}.
\nonumber
\end{eqnarray}
The sum of Eqs. (\ref{e2gt})-(\ref{egt1})-(\ref{egt2}) gives exactly the 
quantity $J_{\a}(t)-J_{\a}^{(\rm S)}$ of Eq. (\ref{DJFNFCO}).

\bibliographystyle{prsty}

\end{document}